\documentclass[nojss]{jss}
\usepackage{amsmath}
\usepackage{amssymb}   
 
\author{Oualid Bada\\University of Bonn \And 
        Dominik Liebl\\Universit\'e libre de Bruxelles}
\title{The \proglang{R}-package \pkg{phtt}: Panel Data Analysis with Heterogeneous Time Trends}

\Plainauthor{Oualid Bada, Dominik Liebl} 
\Plaintitle{The R-package phtt: Panel Data Analysis with Heterogeneous Time Trends} 
\Shorttitle{The \proglang{R}-package \pkg{phtt}} 
 
\Abstract{
  The \proglang{R}-package \pkg{phtt} provides estimation procedures for panel data with large dimensions $n$, $T$, and general forms of unobservable heterogeneous effects. Particularly, the estimation procedures are those of \Citet{Bai2009a} and \Citet{Kneip2009}, which complement one another very well: both models assume the unobservable heterogeneous effects to have a factor structure. \Citet{Kneip2009} considers the case in which the time varying common factors have relatively smooth patterns including strongly positive auto-correlated stationary as well as non-stationary factors, whereas the method of \Citet{Bai2009a} focuses on stochastic bounded factors such as ARMA processes. 
Additionally, the \pkg{phtt} package provides a wide range of dimensionality criteria in order to estimate the number of the unobserved factors simultaneously with the remaining model parameters.  
}
\Plainkeywords{Panel data, unobserved heterogeneity, principal component analysis, factor dimension, interactive fixed effects} 

\Address{
  Oualid Bada\\
  Statistische Abteilung\\
  University of Bonn\\
  Adenauerallee 24-26\\
  53113 Bonn, Germany\\
  E-mail: \email{bada@uni-bonn.de}\\
}

\begin{document}


\pagenumbering{arabic} 
\setcounter{page}{1}

\section{Introduction}\label{intro}
One of the main difficulties and at the same time appealing advantages of panel models is their need to deal with the problem of the unobserved heterogeneity. Classical panel models, such as \textit{fixed effects} or \textit{random effects}, try to model unobserved heterogeneity using dummy variables or structural assumptions on the error term (see, e.g.,~\cite{baltagi2005econometric}). In both cases the unobserved heterogeneity is assumed to remain constant over time within each cross-sectional unit---apart from an eventual common time trend. This assumption might be reasonable for approximating panel data with fairly small temporal dimensions $T$; however, for panel data with large $T$ this assumption becomes very often implausible. 

Nowadays, the availability of panel data with large cross-sectional dimensions $n$ and large time dimensions $T$ has triggered the development of a new class of panel data models. Recent discussions by \Citet{Ahn2006}, \Citet{Pesaran2006}, \Citet{Bai2009a}, \Citet{Bai2009}, and \Citet{Kneip2009} have focused on advanced panel models for which the unobservable individual effects are allowed to have heterogeneous (i.e., individual specific) time trends that can be approximated by a factor structure. The basic form of this new class of panel models can be presented as follows:
\begin{equation}\label{extendMod}
y_{it} = \sum_{j=1}^Px_{itj}\beta_j+ \nu_{it} + \epsilon_{it} \; \textrm{ for } i\in\{1,\ldots,n\} \; \textrm{ and } t\in\{1,\ldots,T\},
\end{equation}
where $y_{it}$ is the dependent variable for each individual $i$ at time $t$, $x_{itj}$ is the $j$th element of the vector of explanatory variables $x_{it}\in\mathbb{R}^P$, and $\epsilon_{it}$ is the idiosyncratic error term. The time-varying individual effects $\nu_{it}\in\mathbb{R}$ of individual $i$ for the time points $t\in\{1,\dots,T\}$ are assumed to be generated by $d$ common time-varying factors. The following two specifications of the time-varying individual effects $\nu_{it}$ are implemented in our \proglang{R} package \pkg{phtt}:
\begin{equation}\label{fs}
\nu_{it}=\left\{
\begin{array}{ccll}
v_{it}   &=&\sum_{l =1}^d \lambda_{il}f_{lt},  &\textrm{for the model of \Citet{Bai2009a},}\\
v_{i}(t) &=&\sum_{l =1}^d \lambda_{il}f_{l}(t),&\textrm{for the model of \Citet{Kneip2009}}.
\end{array}\right.
\end{equation}
Here, $\lambda_{il}$ are unobserved individual loadings parameters, $f_{lt}$ are unobserved common factors for the model of \Citet{Bai2009a}, $f_{l}(t)$ are the unobserved common factors for the model of \Citet{Kneip2009}, and $d$ is the unknown factor dimension. 

Note that the explicit consideration of an intercept in model~\eqref{extendMod} is not necessary but may facilitate interpretation. If $x_{it}$ includes an intercept, the time-varying individual effects $\nu_{it}$ are centered around zero. If $x_{it}$ does not include an intercept, the time-varying individual effects $\nu_{it}$ are centered around the overall mean.

Model~\eqref{extendMod} includes the classical panel data models with additive time-invariant individual effects and common time-specific effects. This model is obtained by choosing $d=2$ with a first common factor $f_{1t}=1$ for all $t\in\{1,\ldots,T\}$ that has individual loadings parameters $\lambda_{i1}$, and a second common factor $f_{2t}$ that has the same loadings parameter $\lambda_{i2}=1$ for all $i\in\{1,\ldots,n\}$. 

An intrinsic problem of factor models lies in the fact that the true factors are only identifiable up to rotation. In order to ensure the uniqueness of these parameters, a number of $d^2$ restrictions are required. The usual normalization conditions are given by
\begin{center}
\begin{tabular}{cll}
(a)& $\frac{1}{T}\sum_{t=1}^T f_{lt}^2 = 1$      & for all $l\in\{1,\ldots,d\}$,\\[1ex]
(b)& $\sum_{t=1}^Tf_{lt}f_{kt} = 0$              & for all $l, k\in\{1, \ldots, d\}$ with $k\neq l$, and\\[1ex]
(c)& $\sum_{i=1}^N\lambda_{il}\lambda_{ik} = 0$  & for all $l, k\in\{1, \ldots, d\}$ with $k\neq l$;
\end{tabular}
\end{center}
see, e.g., \Citet{Bai2009a} and \Citet{Kneip2009}. For the model of \Citet{Kneip2009}, $f_{lt}$ in conditions (a) and (b) has to be replaced by $f_{l}(t)$. As usual in factor models, a certain degree of indeterminacy remains, because the factors can only be determined up to sign changes and different ordering schemes. 

\Citet{Kneip2009} consider the case in which the common factors $f_{l}(t)$ show relatively smooth patterns over time. This includes strongly positive auto-correlated stationary as well as non-stationary factors. The authors propose to approximate the time-varying individual effects $v_{i}(t)$ by smooth nonparametric functions, say, $\vartheta_i(t)$. In this way~\eqref{extendMod} becomes a semi-parametric model and its estimation is done using a two-step estimation procedure, which we explain in more detail in Section~\ref{KSS}. The asymptotic properties of this method rely, however, on independent and identically distributed errors. 

Alternatively, \Citet{Bai2009a} allows for weak forms of heteroskedasticity and dependency in both time and cross-section dimensions and proposes an iterated least squares approach to estimate~\eqref{extendMod} for stationary time-varying individual effects $v_{it}$ such as ARMA processes or non-stationary deterministic trends. However, \Citet{Bai2009a} rules out a large class of non-stationary processes such as stochastic processes with integration. 

Moreover, \Citet{Bai2009a} assumes the factor dimension $d$ to be a known parameter, which is usually not the case. Therefore, the \pkg{phtt} package uses an algorithmic refinement of Bai's method proposed by \Citet{Bada2010} in order to estimate the number of unobserved common factors $d$ jointly with the remaining model parameters; see Section~\ref{Eup} for more details. 

Besides the implementations of the methods proposed by \Citet{Kneip2009}, \Citet{Bai2009a}, and \Citet{Bada2010} the \proglang{R} package \pkg{phtt} comes with a wide range of criteria (16 in total) for estimating the factor dimension $d$. The main functions of the \pkg{phtt} package are given in the following list:
\begin{itemize}
\item \code{KSS()}: Computes the estimators of the model parameters according to the method of \Citet{Kneip2009}; see Section~\ref{KSS}.
\item \code{Eup()}: Computes the estimators of the model parameters according to the method of \Citet{Bai2009a} and \Citet{Bada2010}; see Section~\ref{Eup}.
\item \code{OptDim()}: Allows for a comparison of the estimated factor dimensions $\hat{d}$ obtained from many different (in total $16$) criteria; see Section~\ref{OptDim}.
\item \code{checkSpecif()}: Tests whether to use a classical fixed effects panel model or a panel model with individual effects $\nu_{it}$; see Section~\ref{ADDvsINT}.
\end{itemize}

The functions are provided with the usual \code{print()}-, \code{summary()}-, \code{plot()}-, \code{coef()}- and \code{residuals()}-methods.

Standard methods for estimating models for panel and longitudinal data are also implemented in the \proglang{R} packages \pkg{plm} \citep*{croissant2008panel}, \pkg{nlme} \citep*{nlme}, and \pkg{lme4} \citep*{lme4}; see \Citet{croissant2008panel} for an exhaustive comparison of these packages. Recently, \Citet{Millo2012} published the \proglang{R} package \pkg{splm} for spatial panel data models. The \pkg{phtt} package further extends the toolbox for statisticians and econometricians and provides the possibility of analyzing panel data with large dimensions $n$ and $T$ and considers in the case when the unobserved heterogeneity effects are time-varying. 

To the best of our knowledge, our \pkg{phtt} package \cite{phtt_Rforge} is the first software package that offers the estimation methods of \Citet{Bai2009a} and \Citet{Kneip2009}. Regarding the different dimensionality criteria that can by accessed via the function \code{OptDim()} only those of \Citet{Bai2002} are publicly available as \proglang{MATLAB} codes \citep{MATLAB}  from the homepage of Serena Ng (\url{http://www.columbia.edu/~sn2294/}).

To demonstrate the use of our functions, we re-explore the well known \code{Cigar} dataset, which is frequently used in the literature of panel models. The panel contains the per capita cigarette consumptions of $n = 46$ American states from 1963 to 1992 ($T = 30$) as well as data about the income per capita and cigarette prices (see, e.g., \Citet{baltagi1986estimating} for more details on the dataset). 

We follow \Citet{baltagi2004prediction}, who estimate the following panel model: 
\begin{eqnarray}\label{dypanel}
  \ln({\tt Consumption}_{it})&=&\mu+\beta_1\ln({\tt Price}_{it}) + \beta_2\ln({\tt Income}_{it})+ e_{it}.
\end{eqnarray}
Here, ${\tt Consumption}_{it}$ presents the sales of cigarettes (packs of cigarettes per capita), ${\tt Price}_{it}$ is the average real retail price of cigarettes, and ${\tt Income}_{it}$ is the real disposable income per capita. The index $i\in\{1,\dots,46\}$ denotes the single states and the index $t\in\{1,\dots,30\}$ denotes the year. 


We revisit this model, but allow for a multidimensional factor structure such that
\begin{eqnarray*}
e_{it} &=& \sum_{l=1}^d \lambda_{il} f_{lt} + \epsilon_{it}.
\end{eqnarray*}

\begin{figure}[htbp]
    \centering
    \includegraphics[width=1\textwidth,height=.5\textwidth]{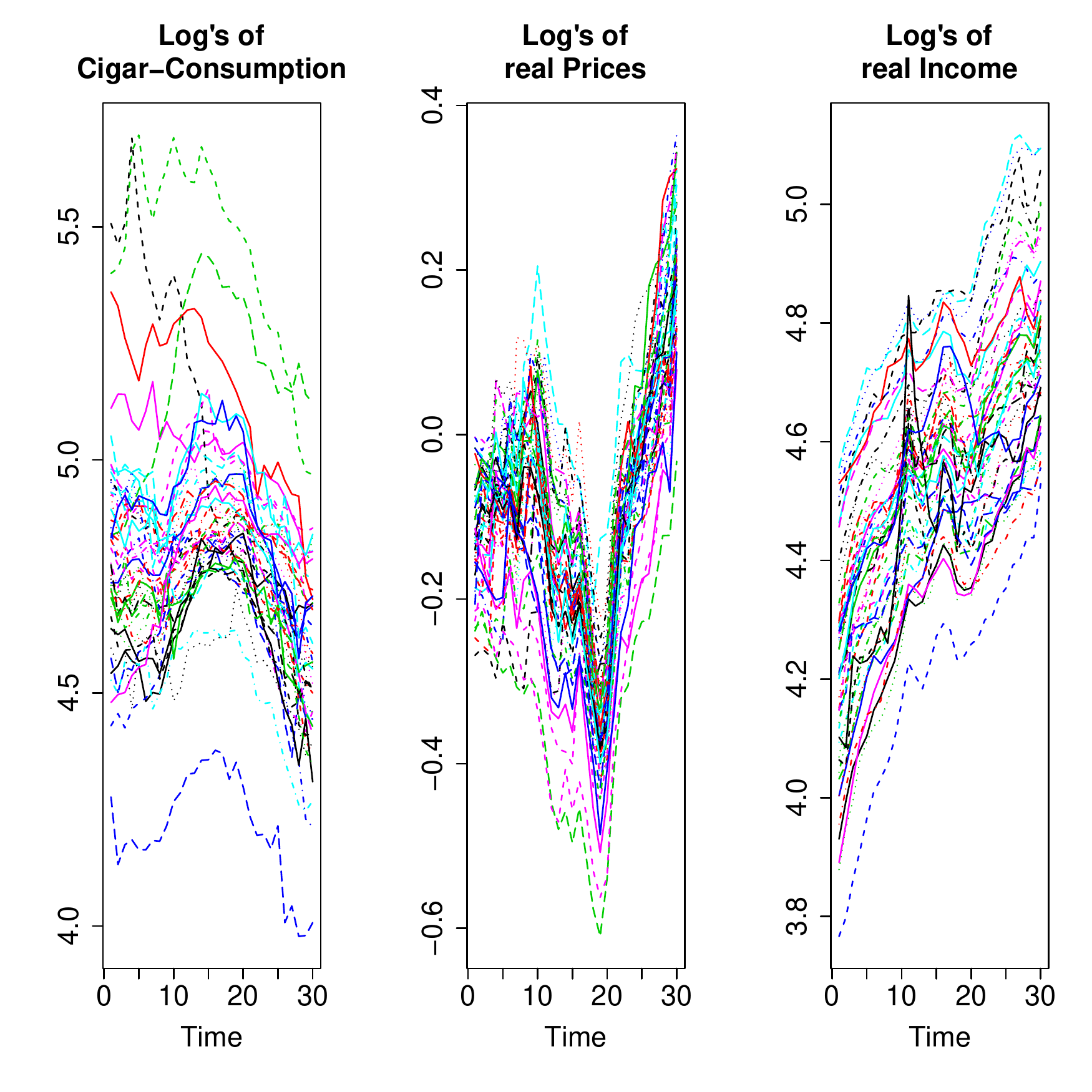} 
    \caption{Plots of the dependent variable $\ln({\tt Consumption}_{it})$ and regressor variables $\ln({\tt Price}_{it})$ and $\ln({\tt Income}_{it})$.}  
  \label{Cigar_Fig} 
\end{figure}

The \code{Cigar} dataset can be obtained from the \pkg{phtt} package using the function \code{data("Cigar")}. The panels of the variables $\ln({\tt Consumption}_{it})$, $\ln({\tt Price}_{it})$, and $\ln({\tt Income}_{it})$ are shown in Figure~\ref{Cigar_Fig}.

Section~\ref{KSS} is devoted to a short introduction of the method of \Citet{Kneip2009}, which is appropriate for relatively smooth common factors $f_l(t)$. Section~\ref{OptDim} presents the usage of the function \code{OptDim()}, which provides access to a wide range of panel dimensionality criteria recently discussed in the literature on factor models. Section~\ref{Eup} deals with the explanation as well as application of the panel method proposed by \Citet{Bai2009a}, which is basically appropriate for stationary and relatively unstructured common factors $f_{lt}$. 

\section{Panel models for heterogeneity in time trends}\label{KSS}
The panel model proposed by \Citet{Kneip2009} can be presented as follows:
\begin{eqnarray}
\label{kss}
y_{it}&=&\sum_{j=1}^Px_{itj}\beta_j+ v_{i}(t)+\epsilon_{it}, 
\end{eqnarray}
where the time-varying individual effects $v_{i}(t)$ are parametrized in terms of common non-parametric basis functions $f_1(t),\dots,f_d(t)$ such that
\begin{eqnarray}\label{kssFS}
v_{i}(t)&=&\sum_{l=1}^d \lambda_{il}f_{l}(t).
\end{eqnarray}

The asymptotic properties of this method rely on second order differences of $v_{i}(t)$, which apply for continuous functions as well as for classical discrete stochastic time series processes such as (S)AR(I)MA processes. Therefore, the functional notation of the time-varying individual effects $v_{i}(t)$ and their underlying common factors $f_1(t),\dots,f_d(t)$ does not restrict them to a purely functional interpretation. The main idea of this approach is to approximate the time series of individual effects $v_{i}(t)$ by smooth functions $\vartheta_i(t)$. 

The estimation approach proposed by \Citet{Kneip2009} relies on a two-step procedure: first, estimates of the common slope parameters $\beta_j$ and the time-varying individual effects $v_{i}(t)$ are obtained semi-parametrically. Second, functional principal component analysis is used to estimate the common factors $f_1(t),\ldots,f_d(t)$, and to re-estimate the time-varying individual effects $v_{i}(t)$ more efficiently. In the following we describe both steps in more detail.

\textbf{Step 1:} The unobserved parameters $\beta_j$ and $v_i(t)$ are estimated by the minimization of
\begin{equation}\label{kssmin}
\sum_{i=1}^n\frac{1}{T}\sum_{t=1}^T\left(y_{it}-\sum_{j=1}^Px_{itj}\beta_j-\vartheta_i(t)\right)^2+\sum_{i=1}^n\kappa\int_{1}^T\frac{1}{T}\left(\vartheta_i^{(m)}(s)\right)^2\,ds,
\end{equation}
over all $\beta_j\in\mathbb{R}$ and all $m$-times continuously differentiable functions $\vartheta_i(t)$, where $\vartheta_i^{(m)}(t)$ denotes the $m$th derivative of the function $\vartheta_i(t)$. A first approximation of $v_i(t)$ is then given by $\tilde{v}_i(t):=\hat\vartheta_i(t)$. Spline theory implies that any solution $\hat{\vartheta}_i(t)$ possesses an expansion in terms of a natural spline basis $z_1(t),\ldots, z_T(t)$ such that $\hat{\vartheta}_i(t) = \sum_{s=1}^T \hat{\zeta}_{is}z_s(t)$; see, e.g., \Citet{de2001practical}. Using the latter expression, we can rewrite~\eqref{kssmin} to formalize the following objective function:
\begin{equation}\label{kssmin2}
S(\beta, \zeta) = \sum_{i=1}^n \left(||Y_i - X_i\beta - Z\zeta_i||^2 + \kappa \zeta_i^{\top}R\zeta_i\right),
\end{equation}
where $Y_i = (y_{i1}, \ldots, y_{iT})^\top$, $X_i = (x_{i1}^\top, \ldots, x_{iT}^\top)^\top$, $\beta=(\beta_1,\dots,\beta_p)^\top$, $\zeta_i=(\zeta_{i1},\dots,\zeta_{iT})^\top$, $Z$ and $R$ are $T \times T$ matrices with elements $\{z_s(t)\}_{s,t = 1, \ldots, T}$ and $\{ \int z^{(m)}_s(t)z^{(m)}_k(t) dt \}_{s,k = 1, \ldots, T}$ respectively. $\kappa$ is a preselected smoothing parameter to control the smoothness of $\hat{\vartheta}_i(t)$. We follow the usual choice of $m = 2$, which leads to cubic smoothing splines. 

In contrast to \Citet{Kneip2009}, we do not specify a common time effect in model \eqref{kss}, but the vector of explanatory variables is allowed to contain an intercept. This means that the time-varying individual effects $v_i(t)$ are not centered around zero for each specific time point $t$, but around a common intercept term. The separate estimation of the common time effect, say $\theta_t$, is also possible with our \pkg{phtt} package; we discuss this in detail in Section \ref{mbshtve}.

The semi-parametric estimators $\hat{\beta}, \hat{\zeta}_i = (\hat{\zeta}_{i1}, \ldots, \hat{\zeta}_{iT})^{\top}$, and $\tilde{v}_i=(\tilde{v}_{i1},\dots,\tilde{v}_{iT})^\top$ can be obtained by minimizing $S(\beta, \zeta)$ over all $\beta\in\mathbb{R}^p$ and $\zeta\in\mathbb{R}^{T\times n}$.

The solutions are given by
	\begin{eqnarray} \label{betaKSS1}
	\hat{\beta} &=& \left(\sum^N_{i = 1} X^{\top}_{i}(I-\mathcal{Z}_{\kappa})X_{i}\right)^{-1} \left(\sum^N_{i = 1}X_i^{\top}(I-\mathcal{Z}_{\kappa})Y_{i} \right),\\
	\hat{\zeta}_i &=& (Z^\top Z+\kappa R)^{-1}Z^\top(Y_i-X_i\hat{\beta}),\textrm{ and}\\
	\tilde{v}_i&=&\mathcal{Z}_{\kappa}\left(Y_i-X_i\hat{\beta} \right),\textrm{ where } \mathcal{Z}_{\kappa} = Z\left(Z^\top Z+\kappa R\right)^{-1}Z^\top.
	\end{eqnarray}

\textbf{Step 2:} The common factors are obtained by the first $d$ eigenvectors $\hat{\gamma}_1, \ldots, \hat{\gamma}_d$ that correspond to the largest eigenvalues $\hat{\rho}_1, \ldots, \hat{\rho}_d$ of the empirical covariance matrix
\begin{equation}\label{Sigmav}
  \hat{\Sigma} = \frac{1}{n} \sum_{i=1}^n\tilde{v}_{i}\tilde{v}^{\top}_i.
\end{equation}	
The estimator of the common factor $f_l(t)$ is then defined by the $l$th scaled eigenvector
\begin{equation}\label{fKSS1}
  \hat{f}_l(t) = \sqrt{T} \hat{\gamma}_{lt} \; \textrm{ for all } l\in\{1, \ldots,d\},
\end{equation}	
where $\hat{\gamma}_{lt}$ is the $t$th element of the eigenvector $\hat{\gamma}_{l}$. The scaling factor $\sqrt{T}$ yields that $\hat{f}_{l}(t)$ satisfies the normalization condition $\frac{1}{T}\sum_{t=1}^T\hat{f}_{l}(t)^2=1$ as listed above in Section~\ref{intro}. The estimates of the individual loadings parameters $\lambda_{il}$ are obtained by ordinary least squares regressions of $\left(Y_{i} - X_{i}\hat{\beta}\right)$ on $\hat{f}_{l}$, where $\hat{f}_{l}=(\hat{f}_{l}(1),\dots,\hat{f}_{l}(T))^{'}$. Recall from conditions (a) and (b) that $\hat{\lambda}_{il}$ can be calculated as follows:
\begin{equation}\label{lambdaKSS1}
\hat{\lambda}_{il} = \frac{1}{T} \hat{f}_{l}^{\top}\left(Y_{i} - X_{i}\hat{\beta}\right).
\end{equation}

A crucial part of the estimation procedure of \Citet{Kneip2009} is the re-estimation of the time-varying individual effects $v_i(t)$ in \textbf{Step 2} by $\hat{v}_i(t):=\sum_{l=1}^d\hat\lambda_l\hat{f}_l(t)$, where the factor dimension $d$ can be determined, e.g., by the sequential testing procedure of \Citet{Kneip2009} or by any other dimensionality criterion; see also Section~\ref{OptDim}. This re-estimation leads to more efficiently estimated time-varying individual effects. 

\Citet{Kneip2009} derive the consistency of the estimators as $n,T\rightarrow\infty$ and show that the asymptotic distribution of common slope estimators is given by $\hat\Sigma_{\beta}^{-1/2}(\hat\beta-\textrm{E}_\epsilon(\hat\beta))\overset{d}{\rightarrow}\mathbf{N}(0,I)$,
where
\begin{equation}\label{asympt.distr2}
\hat\Sigma_{\beta}=\sigma^2\left(\sum_{i=1}^nX_i^\top(I-\mathcal{Z}_\kappa)X_i\right)^{-1}\left(\sum_{i=1}^nX_i^\top(I-\mathcal{Z}_\kappa)^2X_i\right)\left(\sum_{i=1}^nX_i^\top(I-\mathcal{Z}_\kappa)X_i\right)^{-1}.
\end{equation}
A consistent estimator of $\sigma^2$ can be obtained by 
\begin{equation}\label{sig2s2}
\hat{\sigma}^2 = \frac{1}{(n-1)T}\sum_{i=1}^n||Y_i - X_i \hat\beta - \sum_{l=1}^{\hat{d}}\hat\lambda_{i,l} \hat{f}_{l} ||^2. 
\end{equation}

To determine the optimal smoothing parameter $\kappa_{opt}$, \Citet{Kneip2009} propose the following cross validation (CV) criterion: 
\begin{equation}\label{CV}
CV(\kappa) = \sum_{i=1}^n ||Y_i - X_i \hat\beta_{-i} - \sum_{l=1}^d \hat\lambda_{-i,l} \hat{f}_{-i,l}||^2,
\end{equation}
where $\hat\beta_{-i}$, $\hat\lambda_{-i,l}$, and $\hat{f}_{-i,l}$ are estimates of the parameters $\beta$, $\lambda$, and $f_l$ based on the dataset without the $i$th observation. Unfortunately, this criterion is computationally very costly and requires determining the factor dimension $d$ in advance. To overcome this disadvantage, we propose a plug-in smoothing parameter that is discussed in more detail in the following Section~\ref{AlgD}.   

\subsection{Computational details}\label{AlgD}
Theoretically, it is possible to determine $\kappa$ by the CV criterion in~\eqref{CV}; however, cross validation is computationally very costly. Moreover, \Citet{Kneip2009} do not explain how the factor dimension $d$ is to be specified during the optimization process, which is critical since the estimator $\hat{d}$ is influenced by the choice of $\kappa$.

In order to get a quick and effective solution, we propose to determine the smoothing parameter $\kappa$ by generalized cross validation (GCV). However, we cannot apply the classical GCV formulas as proposed, e.g., in \Citet{craven1978smoothing} since we do not know the parameters $\beta$ and $v_{i}(t)$. Our computational algorithm for determining the GCV smoothing parameter $\kappa_{GCV}$ is based on the method of \Citet{Cao2010}, who propose optimizing objective functions of the form~\eqref{kssmin2} by updating the parameters iteratively in a functional hierarchy. Formally, the iteration algorithm can be described as follows:
\begin{enumerate}
\item For given $\kappa$ and ${\beta}$, we optimize~\eqref{kssmin2} with  respect to ${\zeta}_i$ to get
  \begin{equation}
    \label{zetal}
    \hat{\zeta}_i = (Z'Z+\kappa R)^{-1}Z^\top(Y_i-X_i{\beta}).
  \end{equation}
\item By using~\eqref{zetal}, we minimize~\eqref{kssmin2} with respect to $\beta$ to get
  \begin{equation}
    \label{betal}
    \hat{\beta} = \left(\sum^N_{i = 1} X^{\top}_{i}X_{i}\right)^{-1} \left(\sum^N_{i = 1}X_i^\top( Y_{i} - Z\hat{\zeta}_i)\right)
  \end{equation}
\item Once~\eqref{zetal} and~\eqref{betal} are obtained, we optimize the following GCV criterion to calculate $\kappa_{GCV}$: 
  \begin{equation}
    \label{kappal}
\kappa_{GCV} = \arg\min_{\kappa} \frac{1}{\frac{n}{T}tr(I - \mathcal{Z}_{\kappa})^2}\sum_{i=1}^n|| Y_i - X_i \hat{\beta} - \mathcal{Z}_{\kappa}(Y_i - X_i \hat{\beta}) ||^2.
\end{equation}
\end{enumerate}
The program starts with initial estimates of $\beta$ and $\kappa$ and proceeds with steps 1, 2, and 3 in recurrence until convergence of all parameters, where the initial value $\hat\beta_{start}$ is defined in~\eqref{betastart} and the initial value $\kappa_{start}$ is the GCV-smoothing parameter of the residuals $Y_i - X_i \hat\beta_{start}$. 

The advantage of this approach is that the inversion of the $P\times P $ matrix in~\eqref{betal} does not have to be updated during the iteration process. Moreover, the determination of the GCV-minimizer in~\eqref{kappal} can be easily performed in \proglang{R} using the function \code{smooth.spline()}, which calls on a rapid \proglang{C}-routine. 

But note that the GCV smoothing parameter $\kappa_{GCV}$ in~\eqref{kappal} does not explicitly account for the factor structure of the time-varying individual effects $v_{i}(t)$ as formalized in~\eqref{fs}. In fact, given that the assumption of a factor structure is true, the goal shall not be to obtain optimal estimates of $v_i(t)$ but rather to obtain optimal estimates of the common factors $f_{l}(t)$, which implies that the optimal smoothing parameter $\kappa_{opt}$ will be smaller than $\kappa_{GCV}$; see \Citet{Kneip2009}. 

If the goal is to obtain optimal estimates of $f_{l}(t)$, $\kappa_{opt}$ will be used as an upper bound when minimizing the CV criterion \eqref{CV} (via setting the argument \code{CV = TRUE}); which, however, can take some time. Note that, this optimal smoothing parameter $\kappa_{opt}$ depends on the unknown factor dimension $d$. Therefore, we propose to, first, estimate the dimension based on the smoothing parameter $\kappa_{GCV}$ and, second, to use the estimated dimension $\hat{d}$ (via explicitly setting the dimension argument \code{factor.dim}$=\hat{d}$) in order to determine the dimension-specific smoothing parameter $\kappa_{opt}$ (via setting the argument \code{CV = TRUE}).

\subsection{Application }\label{ApplKSS}
This section is devoted to the application of the method of \Citet{Kneip2009} discussed above. The computation of this method is accessible through the function \code{KSS()}, which has the following arguments:

\begin{Schunk}
\begin{Sinput}
R> args(KSS)
\end{Sinput}
\begin{Soutput}
function (formula, additive.effects = c("none", "individual", 
    "time", "twoways"), consult.dim.crit = FALSE, d.max = NULL, 
    sig2.hat = NULL, factor.dim = NULL, level = 0.01, spar = NULL, 
    CV = FALSE, convergence = 1e-06, restrict.mode = c("restrict.factors", 
        "restrict.loadings"), ...) 
NULL
\end{Soutput}
\end{Schunk}

The argument \code{formula} is compatible with the usual \code{R}-specific symbolic designation of the model. The unique specificity here is that the  variables should be defined as $T \times n$ matrices, where $T$ is the temporal dimension and $n$ is the number of the cross-section unites.\footnote{Note that \pkg{phtt} is written for balanced panels. Missing values have to be replaced in a pre-processing step by appropriate imputation methods.}

The argument \code{additive.effects} makes it possible to extend the model~\eqref{kss} for additional additive \code{individual}, \code{time}, or \code{twoways} effects as discussed in Section~\ref{mbshtve}. 

If the logical argument \code{consult.dim.crit} is set to \code{TRUE} all dimensionality criteria discussed in Section~\ref{OptDim} are computed and the user is asked to choose one of their results.

The arguments \code{d.max} and \code{sig2.hat} are required for the computation of some dimensionality criteria discussed in Section~\ref{OptDim}. If their default values are maintained, the function internally computes \code{d.max}$= \left\lfloor \min \{\sqrt{n}, \sqrt{T}\}\right\rfloor$ and \code{sig2.hat} as in~\eqref{sig2s2},  where $\left\lfloor x \right\rfloor$ indicates the integer part of $x$. The argument \code{level} allows to adjust the significance level for the dimensionality testing procedure~\eqref{KSScirt} of \Citet{Kneip2009}; see Section~\ref{OptDim}. 

 \code{CV} is a logical argument. If it is set to \code{TRUE} the cross validation criterion \eqref{CV} of \Citet{Kneip2009} will be computed. In the default case, the function uses the GCV method discussed above in Section \ref{AlgD}.

The factor dimension $d$ can be pre-specified by the argument \code{factor.dim}. Recall from restriction (a) that $\frac{1}{T}\sum_{t=1}^T\hat{f}_l(t)^2=1$. 

Alternatively, it is possible to standardize the individual loadings parameters such that $\frac{1}{n}\sum_{i=1}^n\hat{\lambda}_{il}=1$, which can be done by setting \code{restrict.mode = "restrict.loadings"}.

As an illustration we estimate the Cigarettes model~\eqref{dypanel} introduced in Section~\ref{intro}:
\begin{eqnarray}\label{KSSM1}
  \ln(\texttt{Consumption}_{it})&=&\mu+\beta_1\ln(\texttt{Price}_{it})+ \beta_2\ln(\texttt{Income}_{it}) + e_{it}\\
  \textrm{with}\quad e_{it} &=& \sum_{l=1}^d\lambda_{il}\,f_l(t) + \epsilon_{it},\notag 
\end{eqnarray}
In the following lines of code we load the \code{Cigar} dataset and take logarithms of the three variables, $\texttt{Consumption}_{it}, \texttt{Price}_{it}/\texttt{cpi}_{t}$ and $\texttt{Income}_{it}/\texttt{cpi}_{t}$, where $\texttt{cpi}_{t}$ is the consumer price index. The variables are stored as $T\times n$-matrices. This is necessary, because the \code{formula} argument of the \code{KSS()}-function takes the panel variables as matrices in which the number of rows has to be equal to the temporal dimension $T$ and the number of columns has to be equal to the individual dimension $n$. 

\begin{Schunk}
\begin{Sinput}
R> library("phtt")
R> data("Cigar")
R> N <- 46
R> T <- 30
R> l.Consumption   <- log(matrix(Cigar$sales, T, N))
R> cpi             <- matrix(Cigar$cpi,       T, N)
R> l.Price         <- log(matrix(Cigar$price, T, N)/cpi)
R> l.Income        <- log(matrix(Cigar$ndi,   T, N)/cpi)
\end{Sinput}
\end{Schunk}

The model parameters $\beta_1$, $\beta_2$, the factors $f_l(t)$, the loadings parameters $\lambda_{il}$, and the factor dimension $d$  can be estimated by the  \code{KSS()}-function  with its default arguments. Inferences about the slope parameters can be obtained by using the method \code{summary()}. 

\begin{Schunk}
\begin{Sinput}
R> Cigar.KSS <- KSS(formula = l.Consumption ~ l.Price + l.Income) 
R> (Cigar.KSS.summary <- summary(Cigar.KSS))
\end{Sinput}
\begin{Soutput}
Call:
KSS.default(formula = l.Consumption ~ l.Price + l.Income)

Residuals:
   Min     1Q Median     3Q    Max 
 -0.11  -0.01   0.00   0.01   0.12 


 Slope-Coefficients:
            Estimate  StdErr z.value    Pr(>z)    
(Intercept)   4.0600  0.1770   23.00 < 2.2e-16 ***
l.Price      -0.2600  0.0223  -11.70 < 2.2e-16 ***
l.Income      0.1550  0.0382    4.05  5.17e-05 ***
---
Signif. codes:  0 '***' 0.001 '**' 0.01 '*' 0.05 '.' 0.1 ' ' 1

Additive Effects Type:  none  

Used Dimension of the Unobserved Factors: 6  

Residual standard error: 0.000725 on 921 degrees of freedom 
R-squared: 0.99 
\end{Soutput}
\end{Schunk}

The effects of the log-real prices for cigarettes and the log-real incomes on the log-sales of cigarettes are highly significant and in line with results in the literature. The summary output reports an estimated factor dimension of $\hat{d}=6$. In order to get a visual impression of the six estimated common factors $\hat{f}_1(t),\dots,\hat{f}_6(t)$ and the estimated time-varying individual effects $\hat{v}_1(t),\dots,\hat{v}_n(t)$, we provide a \code{plot()}-method for the \code{KSS}-summary object.

\begin{Schunk}
\begin{Sinput}
R> plot(Cigar.KSS.summary)
\end{Sinput}
\end{Schunk}

\begin{figure}[h]
    \centering
    \includegraphics[width=1\textwidth,height=.5\textwidth]{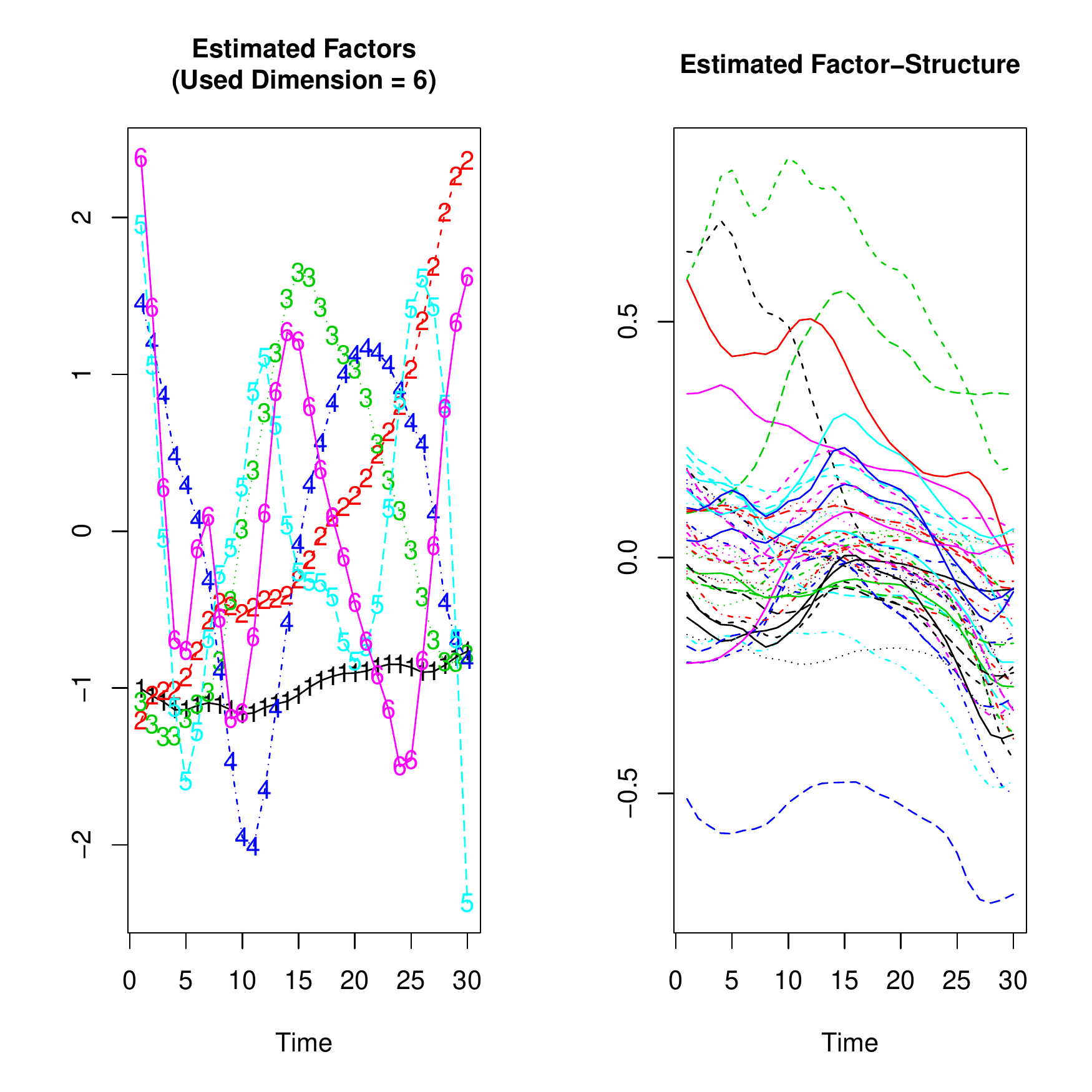} 
    \caption{{\sc Left panel:} Estimated factors $\hat{f}_1(t),\dots,\hat{f}_6(t)$. {\sc Right panel:} Estimated time-varying individual effects $\hat{v}_1(t),\dots,\hat{v}_n(t)$.}  
  \label{KSSM1_Fig} 
\end{figure} 

The left panel of Figure~\ref{KSSM1_Fig} shows the six estimated common factors $\hat{f}_1(t),\dots,\hat{f}_6(t)$ and the right panel of Figure~\ref{KSSM1_Fig} shows the $n=46$ estimated time-varying individual effects $\hat{v}_1(t),\dots,\hat{v}_n(t)$. The common factors are ordered correspondingly to the decreasing sequence of their eigenvalues. Obviously, the first common factor is nearly time-invariant; this suggests extending the model~\eqref{KSSM1} by additive \code{individual} (time-invariante) effects; see Section~\ref{mbshtve} for more details. 

By setting the logical argument \code{consult.dim.crit=TRUE}, the  user can choose from other dimensionality criteria, which are discussed in Section~\ref{OptDim}. Note that the consideration of different factor dimensions $d$ would not alter the results for the slope parameters $\beta$ since the estimation procedure of \Citet{Kneip2009} for the slope parameters $\beta$ does not depend on the dimensionality parameter $d$. 

\section{Panel criteria for selecting the number of factors}\label{OptDim}
In order to estimate the factor dimension $d$, \Citet{Kneip2009} propose a sequential testing procedure based on the following test statistic:
\begin{equation}\label{KSScirt}
  KSS(d) = \frac{n\sum_{r = d+1}^T \hat{\rho}_r - (n-1) \hat{\sigma}^2 tr( \mathcal{Z}_{\kappa} \hat{\mathcal{P}_d}  \mathcal{Z}_{\kappa})}{\hat{\sigma}^2 \sqrt{2N \cdot tr(( \mathcal{Z}_{\kappa} \hat{\mathcal{P}_d}  \mathcal{Z}_{\kappa})^2)}} \stackrel{a}{\sim} N(0,1),
\end{equation}
where $\hat{\mathcal{P}_d} = I - \frac{1}{T}\sum_{l = 1}^d f_lf_l^{\top}$ with $f_l = (f_{l}(1), \ldots, f_{l}(T))^{\top}$, and 
\begin{equation}\label{sig2hkss}
  \hat\sigma^2=\frac{1}{(n-1)tr((I-\mathcal{Z}_\kappa)^2)}\sum_{i=1}^n||(I-\mathcal{Z}_\kappa)(Y_i-X_i\hat\beta)||^2.
\end{equation}
The selection method can be described as follows: choose a significance level $\alpha$ (e.g., $\alpha = 1\%$) and begin with $H_{0}: d = 0$. Test if $KSS(0)\leq z_{1-\alpha}$, where $z_{1-\alpha}$ is the $(1-\alpha)$-quantile of the standard normal distribution. If the null hypothesis can be rejected, go on with $d = 1, 2, 3, \ldots$ until $H_0$ cannot be rejected. Finally, the estimated dimension is then given by the smallest dimension $d$, which leads a rejection of $H_0$.
		
The dimensionality criterion of \Citet{Kneip2009} can be used for stationary as well as non-stationary factors. However, this selection procedure has a tendency to ignore factors that are weakly auto-correlated. As a result, the number of factors can be underestimated.

More robust against this kind of underestimation are the criteria of \Citet{Bai2002}. The basic idea of their approach consists simply of finding a suitable penalty term $g_{nT}$, which countersteers the undesired variance reduction caused by an increasing number of factors $\hat{d}$. Formally, $\hat{d}$ can be obtained by minimizing the following criterion:
\begin{equation}
PC(l) = \frac{1}{nT}\sum^n_{i=1}\sum^T_{t=1}(y_{it} - \hat{y}_{it}(l))^2 + lg_{nT}
\end{equation}
for all $l \in \{1, 2, \ldots\}$, where $\hat{y}_{it}(l)$ is the fitted value for a given factor dimension $l$. To estimate consistently the dimension of stationary factors \Citet{Bai2002} propose specifying $g_{nT}$ by one of the following penalty terms:
\begin{eqnarray}
g^{\textrm{(PC1)}}_{nT} &=& \hat{\sigma}^2\frac{(n+T)}{nT}\log \left(\frac{nT}{n+T}\right), \\
g^{\textrm{(PC2)}}_{nT} &=& \hat{\sigma}^2\frac{(n+T)}{nT}\log(\min\{n, T\}), \\
g^{\textrm{(PC3)}}_{nT} &=& \hat{\sigma}^2 \frac{\log(\min\{n, T\})}{\min\{n, T\}}, \textrm{ and } \\
g^{\textrm{(BIC3)}}_{nT} &=& \hat{\sigma}^2\frac{(n+T - l)}{nT}\log(nT),
\end{eqnarray}
where $\hat{\sigma}^2$ is the sample variance estimator of the residuals $\hat \epsilon_{it}$. The proposed criteria are denoted by PC1, PC2, PC3, and BIC3 respectively. Note that only the first three criteria satisfy the requirements of Theorem 2 in  \Citet{Bai2002}, i.e., $(i) \; g_{nT} \to 0$ and $ (ii) \min \{n, T\}g_{nt} \to \infty$, as $n,T \to \infty$. These conditions ensure consistency of the selection procedure without imposing additional restrictions on the proportional behavior of $n$ and $T$. The requirement $(i)$ is not always fulfilled for BIC3, especially when $n$ is too large relative to $T$ or $T$ is too large relative to $n$ (e.g., $n = \exp(T)$ or $T = \exp(n)$). In practice, BIC3 seems to perform very well, especially when the idiosyncratic errors are cross-correlated.

The variance estimator $\hat{\sigma}^{2}$ can be obtained by 
\begin{equation}
\hat{\sigma}^{2}(d_{max}) = \frac{1}{nT}\sum^n_{i=1}\sum^T_{t=1}(y_{it} - \hat{y}_{it}(d_{max}))^2
\label{sig2h},
\end{equation}
where $d_{max}$ is an arbitrary maximal dimension that is larger than $d$. This kind of variance estimation can, however, be inappropriate in some cases, especially when $\hat \sigma^2(d_{max})$ underestimates the true variance. 
To overcome this problem, \Citet{Bai2002} propose three additional criteria (IC1, IC2, and IC3):
\begin{equation}
IC(l) = \log\left( \frac{1}{nT}\sum^n_{i=1}\sum^T_{t=1}( y_{it} - \hat{y}_{it}(l))^2 \right) + lg_{nT}
\end{equation}
with 
\begin{eqnarray}
g^{\textrm{(IC1)}}_{nT} &=& \frac{(n+T)}{nT}\log(\frac{nT}{n+T}), \\
g^{\textrm{(IC2)}}_{nT} &=& \frac{(n+T)}{nT}\log(\min\{n, T\}), \textrm{ and } \\
g^{\textrm{(IC3)}}_{nT} &=& \frac{\log(\min\{n, T\})}{\min\{n, T\}}.
\end{eqnarray}

In order to improve the finite sample performance of IC1 and IC2, \Citet{Alessi2010} propose to multiply the penalties $g^{\textrm{(IC1)}}_{nT}$ and $g^{\textrm{(IC2)}}_{nT}$ with a positive constant $c$ and apply the calibration strategy of \Citet{Hallin2007}. The choice of $c$ is based on the inspection of the criterion behavior through $J$-different tuples of $n$ and $T$, i.e., $(n_1, T_1), \ldots, (n_J, T_J)$, and for different values of $c$ in a pre-specified grid interval. We denote the refined criteria in our package by ABC.IC1 and ABC.IC2 respectively. Note that such a modification does not affect the asymptotic properties of the dimensionality estimator.

Under similar assumptions, \Citet{Ahn2009} propose selecting $d$ by maximizing the ratio of adjacent eigenvalues (or the ratio of their growth rate). The criteria are referred to as \textit{Eigenvalue Ratio} (ER) and \textit{Growth Ratio} (GR) and defined as following:
\begin{eqnarray}
ER &=& \frac{\hat{\rho}_{l}}{\hat{\rho}_{l+1}} \\
\\
GR &=& \frac{\log\left(\sum^T_{r=l}\hat{\rho}_{r}/\sum^T_{r=l+1}\hat{\rho}_{r} \right)}{\log\left(\sum^T_{r=l+1}\hat{\rho}_{r}/\sum^T_{r=l+2}\hat{\rho}_{r} \right)}.
\end{eqnarray}

Note that the theory of the above dimensionality criteria PC1, PC2, PC3, BIC3, IC1, IC2, IC3, IPC1,IPC2, IPC3, ABC.IC1, ABC.IC2, KSS.C, ER, and GR are developed for stochastically bounded factors. In order to estimate the number of unit root factors, \Citet{Bai2004} proposes the following panel criteria: 
\begin{equation}
IPC(l) = \frac{1}{nT}\sum^n_{i=1}\sum^T_{t=1}(y_{it} - \hat{y}_{it}(l))^2 + lg_{nT},
\end{equation}
where
\begin{eqnarray}
g^{\textrm{(IPC1)}}_{nT} &=& \hat{\sigma}^2\frac{\log(\log(T))}{T}\frac{(n+T)}{nT}\log\left(\frac{nT}{n+T}\right), \\
g^{\textrm{(IPC2)}}_{nT} &=& \hat{\sigma}^2\frac{\log(\log(T))}{T}\frac{(n+T)}{nT}\log(\min\{n, T\}), \textrm{ and} \\
g^{\textrm{(IPC3)}}_{nT} &=& \hat{\sigma}^2\frac{\log(\log(T))}{T} \frac{(n+T - l)}{nT}\log(nT).
\end{eqnarray}

Alternatively, \Citet{Onatski2009} has introduced a threshold approach based on the empirical distribution of the sample covariance eigenvalues, which can be used for both stationary and non-stationary factors. The estimated dimension is obtained by
\[
\hat{d} = \max \{l \leq d_{max}: \hat{\rho}_{l} - \hat{\rho}_{l-1} \geq \delta\},
\]
where $\delta$ is a positive threshold, estimated iteratively from the data. We refer to this criterion as ED, which stands for Eigenvalue Differences.

\subsection{Application}
The dimensionality criteria introduced above are implemented in the function \code{OptDim()}, which has the following arguments:
\begin{Schunk}
\begin{Sinput}
R> args(OptDim)
\end{Sinput}
\begin{Soutput}
function (Obj, criteria = c("PC1", "PC2", "PC3", "BIC3", "IC1", 
    "IC2", "IC3", "IPC1", "IPC2", "IPC3", "ABC.IC1", "ABC.IC2", 
    "KSS.C", "ED", "ER", "GR"), standardize = FALSE, d.max, sig2.hat, 
    spar, level = 0.01, c.grid = seq(0, 5, length.out = 128), 
    T.seq, n.seq) 
NULL
\end{Soutput}
\end{Schunk}

The desired criteria can be selected by one or several of the following character variables: \code{"KSS.C"}, \code{"PC1"}, \code{"PC2"}, \code{"PC3"}, \code{"BIC2"}, \code{"IC1"}, \code{"IC2"} , \code{"IC3"}, \code{"ABC.IC1"}, \code{"ABC.IC2"}, \code{"ER"},  \code{"GR"}, \code{"IPC1"}, \code{"IPC2"}, \code{"IPC3"}, and \code{"ED"}. The default significance level used for the \code{"KSS"}-criterion is \code{level = 0.01}. The values of $d_{max}$ and $\hat{\sigma}^2$ can be specified externally by the arguments \code{d.max} and \code{sig2.hat}. By default, \code{d.max} is computed internally as \code{d.max}$= \left\lfloor \min \{\sqrt{n}, \sqrt{T}\}\right\rfloor$ and \code{sig2.hat} as in (\ref{sig2hkss}) and (\ref{sig2h}). The arguments \code{"c.grid"},  \code{"T.seq"}, and \code{"n.seq"} are required for computing \code{"ABC.IC1"} and \code{"ABC.IC2"}. The grid interval of the calibration parameter can be externally specified with \code{"c.grid"}. The $J$-Tuples, $(n_1,T_1), \ldots, (n_J,T_J)$, can be specified by using appropriate vectors in \code{"T.seq"}, and \code{"n.seq"}. If these two arguments are left unspecified, the function constructs internally the following sequences:  $T-C, T-C+1, \dots , T$, and $n-C, n-C+1, \dots , n$, for $C=min{\sqrt{n},\sqrt{T},30}$. Alternatively, the user can specify only the length of the sequences by giving appropriate integers to  the arguments \code{"T.seq"}, and \code{"n.seq"}, to control for $C$. 

The input variable can be standardized by choosing \code{standardize = TRUE}. In this case, the calculation of the eigenvalues is based on the correlation matrix instead of the covariance matrix for all criteria.

As an illustration, imagine that we are interested in the estimation of the factor dimension of the variable $\ln(\texttt{Consumption}_{it})$ with the dimensionality criterion \code{"PC1"}. The function \code{OptDim()} requires a $T \times n$ matrix as input variable.
\begin{Schunk}
\begin{Sinput}
R> OptDim(Obj = l.Consumption, criteria = "PC1")
\end{Sinput}
\begin{Soutput}
Call: OptDim.default(Obj = l.Consumption, criteria = "PC1")

---------
Criterion of Bai and Ng (2002):

  PC1
    5
\end{Soutput}
\end{Schunk}

\code{OptDim()} offers the possibility of comparing the result of different selection procedures by giving the corresponding criteria to the argument \code{criteria}. If the argument \code{criteria} is left unspecified, \code{OptDim()} automatically compares all $16$ procedures. 

\begin{Schunk}
\begin{Sinput}
R> (OptDim.obj <- OptDim(Obj = l.Consumption, criteria = c("PC3",  "ER",  
+                       "GR", "IPC1", "IPC2", "IPC3"), standardize = TRUE))
\end{Sinput}
\begin{Soutput}
Call: OptDim.default(Obj = l.Consumption, criteria = c("PC3", "ER", 
    "GR", "IPC1", "IPC2", "IPC3"), standardize = TRUE)

---------
Criterion of Bai and Ng (2002):

  PC3
    5

--------
Criteria of Ahn and Horenstein (2013):

  ER GR
   3  3

---------
Criteria of Bai (2004):

  IPC1 IPC2 IPC3
     3    3    2
\end{Soutput}
\end{Schunk}

\begin{figure}[h]
    \centering
    \includegraphics[width=0.65\textwidth,height= 0.65\textwidth]{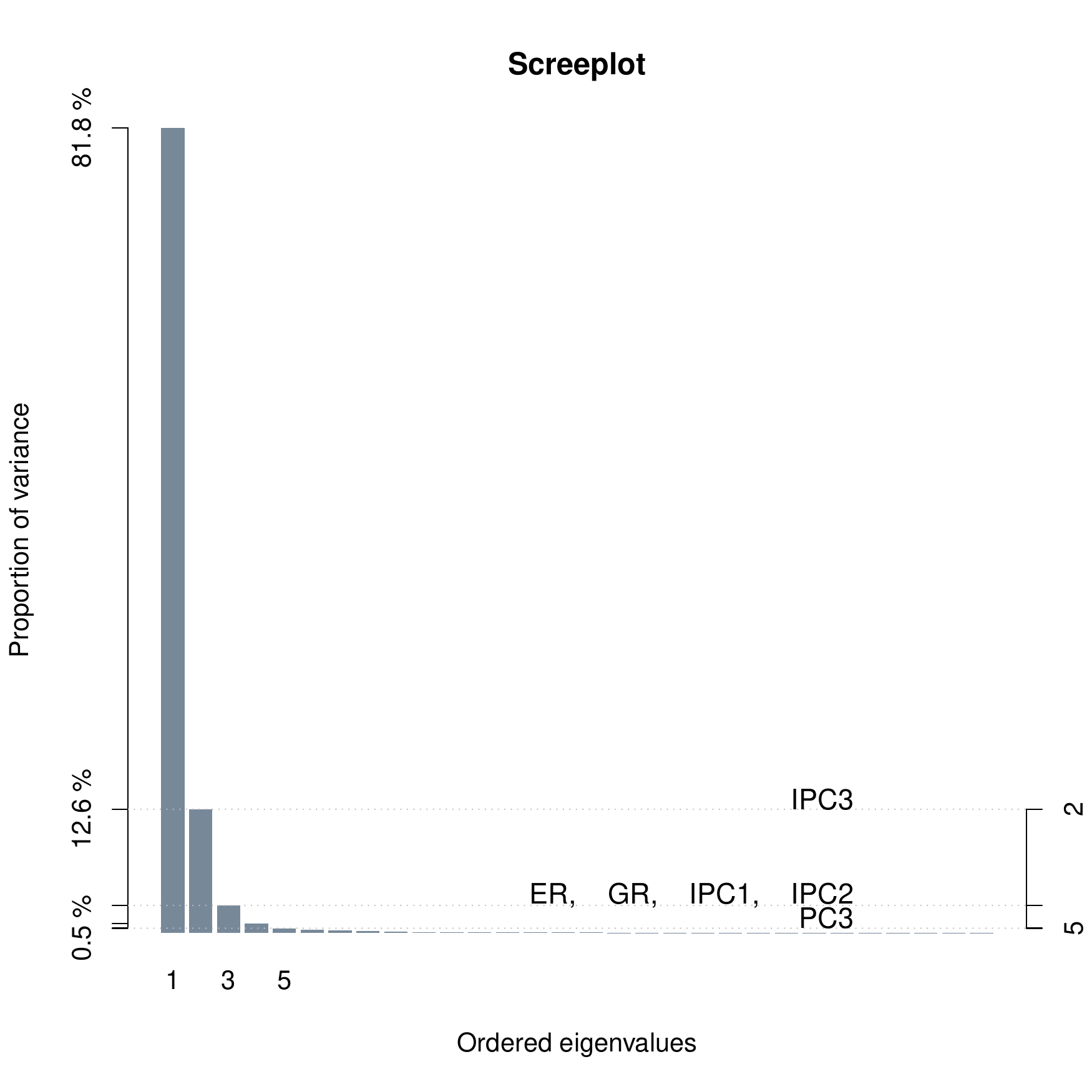} 
    \caption{Scree plot produced by the \code{plot()}-method for \code{OptDim}-objects. Most of the dimensionality criteria (ER, GR, IPC1 and IPC2) suggest using the dimension $\hat{d}=3$.}  
  \label{OptDim_Fig} 
\end{figure}

In order to help users to choose the most appropriate dimensionality criterion for the data, \code{OptDim}-objects are provided with a \code{plot()}-method. This method displays, in descending order, the magnitude of the eigenvalues in percentage of the total variance and indicates where the selected criteria detect the dimension; see Figure~\ref{OptDim_Fig}.

\begin{Schunk}
\begin{Sinput}
R> plot(OptDim.obj) 
\end{Sinput}
\end{Schunk}

We, now, come back to the \code{KSS}- function, which offers an additional way to compare the results of all dimensionality criteria and to select one of them: If the \code{KSS()}-argument \code{consult.dim = TRUE}, the results of the dimensionality criteria are printed on the console of \proglang{R} and the user is asked to choose one of the results.  

\begin{Schunk}
\begin{Sinput}
R> KSS(formula = l.Consumption ~ -1 + l.Price + l.Income, consult.dim = TRUE) 
\end{Sinput}
\end{Schunk}

\begin{CodeChunk}
\begin{CodeOutput}
  -----------------------------------------------------------
Results of Dimension-Estimations

-Bai and Ng (2002):
 PC1  PC2  PC3 BIC3  IC1  IC2  IC3 
   5    5    5    4    5    5    5 


-Bai (2004):
IPC1 IPC2 IPC3 
   3    3    2 


-Alessi et al. (2010):
ABC.IC1 ABC.IC2 
      3       3 

-Kneip et al. (2012):
 KSS.C 
     6 

-Onatski (2009):
 ED 
  3 

-Ahn and Horenstein (2013):
 ER  GR 
  3   6 

-----------------------------------------------------------
Please, choose one of the proposed integers: 
\end{CodeOutput}
\end{CodeChunk}
After entering a number of factors, e.g., $6$ we get the following feedback:
\begin{CodeChunk}
\begin{CodeOutput}
Used dimension of unobs. factor structure is: 6 
-----------------------------------------------------------
\end{CodeOutput}
\end{CodeChunk}
Note that the maximum number of factors that can be given, cannot exceed the highest estimated factor dimension (here maximal dimension would be 6). A higher dimension can be chosen using the argument \code{factor.dim}.

\section{Panel models with stochastically bounded factors}\label{Eup}
The panel model proposed by \Citet{Bai2009a} can be presented as follows:
\begin{equation}
\label{bai2009}
y_{it} =\sum_{j=1}^Px_{itj}\beta_j+ v_{it}+\epsilon_{it}, 
\end{equation}
where 
\begin{equation}\label{baiFS}
v_{it} = \sum_{l=1}^d \lambda_{il}f_{lt}.
\end{equation}
Combining (\ref{bai2009}) with (\ref{baiFS}) and writing the model in matrix notation we get
\begin{equation}\label{modvectorpres}
Y_i = X_i \beta + F \Lambda_i^{\top} + \epsilon_i,
\end{equation}
where $Y_i = (y_{i1}, \ldots, y_{iT})^{\top}$, $X_i = (x_{i1}^{\top}, \ldots, x_{iT}^{\top})^{\top}$, $\epsilon_i = (\epsilon_{i1}, \ldots, \epsilon_{iT})^{\top}$, $\Lambda_i = (\lambda_{1}, \ldots, \lambda_{n})^{\top}$ and $F = (f_1, \ldots, f_T)^{\top}$ with $\lambda_i = (\lambda_{i1}, \ldots, \lambda_{id})$, $f_t = (f_{1t}, \ldots, f_{dt})$, and $\epsilon_i = (\epsilon_{i1}, \ldots, \epsilon_{iT})^{\top}$. 

The asymptotic properties of Bai's method rely, among others, on the following assumption:
\begin{equation}\label{stasiocond}
\frac{1}{T}F^{\top}F \stackrel{p}{\rightarrow} \Sigma_F, \; \textrm{ as }\; T \rightarrow \infty,
\end{equation}
where $\Sigma_F$ is a fixed positive definite $d\times d$  matrix. This allows for the factors to follow a deterministic time trend such as $f_t = t/T$ or to be stationary dynamic processes such that $f_t = \sum_{j=1}^\infty C_j e_{t-j}$, where $e_t$ are i.i.d.~zero mean stochastic components. It is, however, important to note that such an assumption rules out a large class of non-stationary factors such as I($p$) processes with $p \geq1$.

\subsection{Model with known number of factors $d$}
\Citet{Bai2009a} proposes to estimate the model parameters $\beta, F$ and $ \Lambda_i$ by minimizing the following least squares objective function: 
	\begin{equation}
		S(\beta, F, \Lambda_i) = \sum^n_i{||Y_i - X_i\beta - F \Lambda_i^{\top}||^2}.
	\label{of}
	\end{equation}
	For each given $F$, the OLS estimator of $\beta$ can be obtained by
	\begin{equation}\label{betabai}
	\hat{\beta}(F) = \left(\sum^{n}_{i=1}X^\top_{i}{\mathcal{P}_d}X_{i}\right)^{-1} \left(\sum^{n}_{i=1}X^\top_{i} \mathcal{P}_d Y_i \right)
	\end{equation}
	where $\mathcal{P}_d = I - F(F^{\top}F)^{-1}F^{\top} = I - FF^{\top}/T $. If $\beta$ is known, $F$ can be estimated by using the first $d$ eigenvectors $\hat{\gamma} = (\hat{\gamma}_1, \ldots, \hat{\gamma}_d)$ corresponding to the first $d$ eigenvalues of the empirical covariance matrix $\hat{\Sigma} = (nT)^{-1}\sum_{i=1}^n w_iw_i^{\top}$, where $w_i = Y_i - X_i \beta$. That is,
	\begin{equation*}	
	\hat{F}(\beta)  =  \sqrt{T}\hat{\gamma}.
	\end{equation*}
The idea of \Citet{Bai2009a} is to start with initial values for $\beta$ or $F$ and calculate the estimators iteratively.  The method requires, however, the factor dimension $d$ to be known, which is usually not the case in empirical applications. 

A feasible estimator of \eqref{betabai} can be obtained by using an arbitrary large dimension $d_{max}$ greater than $d$. The factor dimension can be estimated subsequently by using the criteria of \Citet{Bai2002} to the remainder term $Y_{i} = X_{i}\hat\beta(\hat F (d_{max}))$, as suggested by \Citet{Bai2009a}. This strategy can lead, however, to inefficient estimation and spurious interpretation of $\beta$ due to over-parameterization. 

\subsection{Model with unknown number of factors $d$}\label{Eup2}
In order to estimate $d$ jointly with $\beta, F$, and $ \Lambda_i$, \Citet{Bada2010} propose to integrate a penalty term into the objective function to be globally optimized.  In this case, the optimization criterion can be defined as a penalized least squares objective function of the form:
\begin{equation}
  S(\beta, F, \Lambda_i, l) = \sum^N_i{||Y_i - X_i\beta - F \Lambda_i^{\top}||^2} + l g_{nT}
  \label{ofp}
\end{equation}
The role of the additional term $lg_{nT}$ is to pick up the dimension $\hat{d}$, of the unobserved factor structure. The penalty $g_{nT}$ can be chosen according to \Citet{Bai2002}. The estimation algorithm is based on the parameter cascading strategy of \Citet{Cao2010}, which in this case can be described as follows:
\begin{enumerate}
\item Minimizing (\ref{ofp}) with respect to $\Lambda_i$ for each given $ \beta, F$ and $d$, we get
		\begin{equation}\label{lambda}
		\hat{\Lambda}^{\top}_i(\beta, F, d) = F^\top\left(Y_i - X_i\beta\right)/T.							 
		\end{equation}
\item Introducing (\ref{lambda}) in (\ref{ofp}) and minimizing with respect to $F$ for each given $\beta$ and $d$, we get
		\begin{eqnarray}\label{theta} 
		\hat{F}(\beta, d)  =  \sqrt{T}\hat{\gamma}(\beta, d),
		\end{eqnarray}
		where $\hat{\gamma}(\beta, d)$ is a $T \times d$ matrix that contains the first $d$ eigenvectors corresponding to the first $d$ eigenvalues $\rho_1, \ldots, \rho_d$ of the covariance matrix $\hat{\Sigma} = (nT)^{-1}\sum_{i=1}^n w_iw_i^{\top}$ with $w_i = Y_i - X_i \beta$.
\item Reintegrating (\ref{theta}) and (\ref{lambda}) in (\ref{ofp}) and minimizing with respect to $\beta$ for each given $d$, we get
		\begin{equation}
		\hat{\beta}(d) = \left(\sum^{N}_{i=1}X^\top_{i}X_{i}\right)^{-1} \left(\sum^{N}_{i=1}X^\top_{i}\left(Y_i -\hat{F}\hat{\Lambda}^{\top}_i(\hat{\beta}, d) \right)\right).
		\label{cup-beta}
		\end{equation} 
\item Optimizing~\eqref{ofp} with respect to $l$ given the results in~\eqref{lambda},~\eqref{theta}, and~\eqref{cup-beta} allows us to select $\hat{d}$ as
		\begin{equation*}
		\hat{d} = \text{argmin}_{l} \sum^N_i ||Y_i - X_i\hat{\beta} - \hat{F} \hat{\Lambda}_i^{\top}||^2 + l g_{nT}, \quad \textrm{ for all } l \in \{ 0,1, \ldots , d_{max}\}.
 		\end{equation*}
\end{enumerate}		

The final estimators are obtained by alternating between an inner iteration to optimize $\hat{\beta}(d), \hat{F}(d)$, and $\hat{\Lambda}_i(d)$ for each given $d$ and an outer iteration to select the dimension $\hat{d}$. The updating process is repeated in its entirety till the convergence of all the parameters. This is why the estimators are called \textsl{entirely updated estimators} (Eup). In order to avoid over-estimation, \Citet{Bada2010}  propose to re-scale $g_{nT}$ in each iteration stage with $\hat \sigma^2 = \sum^N_i ||Y_i - X_i\hat{\beta} - \hat{F} \hat{\Lambda}_i^{\top}||^2$ in stead of $\hat \sigma^2 (d_{max})$. Simulations show that such a calibration can improve the finite sample properties of the estimation method.
 		
It is notable that the objective functions (\ref{ofp}) and (\ref{of}) are not globally convex. There is no guarantee that the iteration algorithm  converges to the global optimum. Therefore, it is important to choose reasonable starting values $\hat{d}_{start}$ and $\hat\beta_{start}$. We propose to select a large dimension $d_{max}$ and to start the iteration with the following estimate of $\beta$: 
\begin{equation}\label{betastart}
  {\hat\beta}_{start} = \left(\sum^{n}_{i=1}X^\top_{i}(I - {G}{G}^{\top})X_{i}\right)^{-1} \left(\sum^{n}_{i=1}X^\top_{i} (I - {G}{G}^{\top}) Y_i \right),
\end{equation} 
where ${G}$ is the $T \times d_{max}$ matrix of the eigenvectors corresponding to the first $d_{max}$ eigenvalues of the augmented covariance matrix 
\begin{equation*}
  \Gamma^{Aug} = \frac{1}{nT} \sum_{i = 1}^n (Y_i, X_i)(Y_i^{\top}, X_i^{\top})^{\top}.
\end{equation*} 

The intuition behind these starting estimates relies on the fact that the unobserved factors cannot escape from the space spanned by the eigenvectors $G$. The projection of $X_i$ on the orthogonal complement of $G$ in~\eqref{betastart} eliminates the effect of a possible correlation between the observed regressors and unobserved factors, which can heavily distort the value of $\beta^{0}$ if it is neglected. \Citet{greenaway2010asymptotic} give conditions under which~\eqref{betastart} is a consistent estimator of $\beta$. In order to avoid miss-specifying the model through identifying factors that only exist in $X_i$ and not $Y_i$, \Citet{Bada2010} recommend to under-scale the starting common factors $G_l$ that are highly correlated with $X_i$.

According to \Citet{Bai2009a}, the asymptotic distribution of the slope estimator $\hat{\beta}(d)$ for known $d$ is given by 
\begin{equation*}
\sqrt{nT}(\hat{\beta}(d) - \beta) \stackrel{a}{\sim} N(0 , D_0^{-1}D_Z D_0^{-1}),
\end{equation*}
where $D_0 = \textrm{plim} \frac{1}{nT} \sum_{i=1}^n \sum_{t=1}^T  Z_{it}^{\top} Z_{it}$ with $Z_{i} = (Z_{i1}, \ldots, Z_{iT})^{\top} = \mathcal{P}_d X_i - \frac{1}{n} \sum_{k=1}^n \mathcal{P}_d X_i a_{ik}$ and $ a_{ik} = \Lambda_i (\frac{1}{n} \sum_{i=1}^n \Lambda_i^{\top}\Lambda_i)^{-1} \Lambda_k^{\top}$, and 
\begin{enumerate}
\item[Case 1.] $D_Z =  D_0^{-1} \sigma^2$ if the errors are i.i.d.~with zero mean and variance $\sigma^2$,
\item[Case 2.] $D_Z = \textrm{plim} \frac{1}{nT} \sum_{i=1}^n \sigma^2_i \sum_{t=1}^T  Z_{it}^{\top} Z_{it}$, where $\sigma^2_i = E(\epsilon^2_{it})$ with $E(\epsilon_{it}) = 0$, if cross-section heteroskedasticity exists and  $n/T \to 0$,
\item[Case 3.] $D_Z = \textrm{plim} \frac{1}{nT} \sum_{i=1}^n  \sum_{j=1}^n \omega_{ij} \sum_{t=1}^T  Z_{it}^{\top} Z_{jt}$, where $\omega_{ij} = E(\epsilon_{it} \epsilon_{jt})$ with $E(\epsilon_{it}) = 0$, if cross-section correlation and heteroskedasticity exist and $n/T \to 0$,
\item[Case 4.] $D_Z = \textrm{plim} \frac{1}{nT} \sum_{t=1}^T \sigma^2_t \sum_{i=1}^n  Z_{it}^{\top} Z_{it}$, where $\sigma^2_t = E(\epsilon^2_{it})$ with $E(\epsilon_{it}) = 0$, if heteroskedasticity in the time dimension exists and $T/n \to 0$,
\item[Case 5.] $D_Z = \textrm{plim} \frac{1}{nT} \sum_{t=1}^T \sum_{s=1}^T \rho(t,s) \sum_{i=1}^n  Z_{it}^{\top} Z_{is}$, where $\rho(t,s) = E(\epsilon_{it}\epsilon_{is})$ with $E(\epsilon_{it}) = 0$ , if correlation and heteroskedasticity in the time dimension exist and $T/n \to 0$, and
\item[Case 6.] $D_Z = \textrm{plim} \frac{1}{nT} \sum_{t=1}^T \sum_{i=1}^n \sigma^2_{it}  Z_{it}^{\top} Z_{is}$, where $\sigma^2_{it} = E(\epsilon_{it}^2)$ with $E(\epsilon_{it}) = 0$, if heteroskedasticity in both time and cross-section dimensions exists with $T/n^2 \to 0$ and $n/T^2 \to 0$.
\end{enumerate}

In presence of correlation and heteroskedasticity in panels with proportional dimensions $n$ and  $T$,  i.e., $n/T \to c > 0$, the asymptotic distribution of $\hat{\beta}(d)$ will be not centered at zero. This can lead to false inference when using the usual test statistics such as $t$- and $\chi^2$-statistic. To overcome this problem, \Citet{Bai2009a} propose to estimate the asymptotic bias and correct the estimator as follows: 
\begin{equation}
\hat \beta^*(d) = \hat \beta(d) - \frac{1}{n}\hat B - \frac{1}{T}\hat C
\end{equation}
where $\hat B$ and $\hat C$ are the estimators of
$$
\begin{array}{lcl}
B &=& - \left(\frac{1}{nT} \sum_{i=1}^n \sum_{t=1}^T  Z_{it}^{\top} Z_{it}\right)^{-1} \frac{1}{n} \sum_{i=1}^n \sum_{k=1}^n(X_i - V_i)^{\top}F \left( F^{\top} F\right)^{-1} \\
C &=&  - \left(\frac{1}{nT} \sum_{i=1}^n \sum_{t=1}^T  Z_{it}^{\top} Z_{it}\right)^{-1} \frac{1}{n} \sum_{i=1}^n X_i^{\top} M_{F}\Omega F \left( F^{\top} F\right)^{-1} \left(\sum_{k=1}^n \Lambda_k^{\top} \Lambda_k \right)^{-1}\Lambda_i^{\top}
\end{array}
$$ 
respectively. Here, $V_i = \frac{1}{n} \sum_{j=1}^{n} a_{ij}X_{j}$,  $\Omega = \frac{1}{n}\sum_{k=1}^{n} \Omega_k$ and
\begin{enumerate}
\item[Case 7.] $\Omega_k$ is a $T \times T$ diagonal matrix with elements $\omega_{kt} = E(\epsilon_{kt}^2)$ if heteroskedasticity in both time and cross-section dimensions exist and $n/T \to c > 0$ and,
\item[Case 8.] $\Omega_k$ is a $T \times T$ matrix with elements $\Omega_{k,ts} = E(\epsilon_{kt} \epsilon_{ks})$ if correlation and heteroskedasticity in both time and cross-section dimensions exist and  $n/T \to c > 0$. 
\end{enumerate}

In a similar context, \Citet{Bada2010} prove that estimating $d$ with the remaining model parameters does not affect the asymptotic properties of $\hat{\beta}({d})$. The asymptotic distribution of $\hat{\beta} = \hat\beta(\hat d)$ is given by 
$$
\sqrt{nT}(\hat{\beta} - \beta) \stackrel{a}{\sim} N(0 , D_0^{-1}D_Z D_0^{-1}) 
$$
under Cases 1-6, and 
$$
\sqrt{nT}(\hat{\beta}^* - \beta) \stackrel{a}{\sim} N(0 , D_0^{-1}D_Z D_0^{-1}) 
$$ 
under Cases 7-8, where $\hat{\beta}^* =  \hat{\beta}^*(\hat d)$.

The asymptotic variance of $\hat{\beta}$ and the bias terms $B$ and $C$ can be estimated by replacing $F$, $ \Lambda_i$, $Z_{it}$, and $\epsilon_{it}$ with $\hat F$, $\hat \Lambda_i$, $\hat Z_{it}$, and $\hat \epsilon_{it}$ respectively. 

In presence of serial correlation (cases 5 and 8), consistent estimators for $D_Z$ and $C$ can be obtained by using the usual heteroskedasticity and autocorrelation (HAC) robust limiting covariance. In presence of cross-section correlation (case 3), $D_Z$ is estimated by  $\hat D_Z = \frac{1}{mT} \sum_{i=1}^m\sum_{j=1}^m \sum_{t=1}^T \hat Z_{it}^{\top}\hat Z_{jt} \hat\epsilon_{it}\hat\epsilon_{jt}$, where $m = \sqrt{n}$. If both cross-section and serial correlation exist (case 8), we estimate the long-run covariance of $\frac{1}{\sqrt{m}}\sum_{j=1}^m \hat Z_{it}\hat\epsilon_{it}$. 

\subsection{Application}
The above described methods are implemented in the function \code{Eup()}, which takes the following arguments: 
\begin{Schunk}
\begin{Sinput}
R> args(Eup)
\end{Sinput}
\begin{Soutput}
function (formula, additive.effects = c("none", "individual", 
    "time", "twoways"), dim.criterion = c("PC1", "PC2", "PC3", 
    "BIC3", "IC1", "IC2", "IC3", "IPC1", "IPC2", "IPC3"), d.max = NULL, 
    sig2.hat = NULL, factor.dim = NULL, double.iteration = TRUE, 
    start.beta = NULL, max.iteration = 500, convergence = 1e-06, 
    restrict.mode = c("restrict.factors", "restrict.loadings"), 
    ...) 
NULL
\end{Soutput}
\end{Schunk}

The arguments \code{additive.effects}, \code{d.max}, \code{sig2.hat}, and \code{restrict.mode} have the same roles as in \code{KSS()}; see Section \ref{ApplKSS}. The argument \code{dim.criterion} specifies the dimensionality criterion to be used if \code{factor.dim} is left unspecified and defaults to \code{dim.criterion = "PC1"}. 

Setting the argument \code{double.iteration=FALSE} may speed up computations, because the updates of $\hat{d}$ will be done simultaneously with $\hat{F}$ without waiting for their inner convergences. However, in this case, the convergence of the parameters is less stable than in the default setting. 

The argument \code{start.beta} allows us to give a vector of starting values for the slope parameters $\beta_{start}$. The maximal number of iteration and the convergence condition can be controlled by \code{max.iteration} and \code{convergence}. 

In our application, we take first-order differences of the observed time series. This is because some factors show temporal trends, which can violate the stationarity condition~\eqref{stasiocond}; see Figure~\ref{KSSM1_Fig}. We consider the following modified cigarettes model: 
\begin{eqnarray*}
\triangledown \ln(\texttt{Consumption}_{it})&=&\beta_1 \triangledown  \ln(\texttt{Price}_{it})+ \beta_2 \triangledown \ln(\texttt{Income}_{it}) +  e_{it},\\
\textrm{with}\quad e_{it} &=&\sum_{l=1}^d\lambda_{il}f_{lt}+ \epsilon_{it}\notag,
\end{eqnarray*}
where $\triangledown x_t = x_t - x_{t-1}$. In order to avoid notational mess, we use the same notation for the unobserved time-varying individual effects $v_{it}= \sum_{l=1}^d\lambda_{il}f_{lt}$ as above in~\eqref{KSSM1}. The $\triangledown$-transformation can be easily performed in \proglang{R} using the standard \code{diff()}-function as follows:
\begin{Schunk}
\begin{Sinput}
R> d.l.Consumption  <- diff(l.Consumption)
R> d.l.Price        <- diff(l.Price)
R> d.l.Income       <- diff(l.Income)
\end{Sinput}
\end{Schunk}

As previously mentioned for the \code{KSS()}-function, the \code{formula} argument of the \code{Eup()}-function takes balanced panel variables as $T\times n$ dimensional matrices, where the number of rows has to be equal to the temporal dimension $T$ and the number of columns has to be equal to the individual dimension $n$. 
\begin{Schunk}
\begin{Sinput}
R> (Cigar.Eup <- Eup(d.l.Consumption ~  -1 + d.l.Price + d.l.Income, 
+                   dim.criterion = "PC3"))
\end{Sinput}
\begin{Soutput}
Call:
Eup.default(formula = d.l.Consumption ~ -1 + d.l.Price + d.l.Income, 
    dim.criterion = "PC3")

Coeff(s) of the Observed Regressor(s) :

  d.l.Price d.l.Income
 -0.3140143   0.159392

Additive Effects Type:  none  

Dimension of the Unobserved Factors: 5  

Number of iterations: 55 
\end{Soutput}
\end{Schunk}

Inferences about the slope parameters can be obtained by using the method \code{summary()}. The type of correlation and heteroskedasticity in the idiosyncratic errors can be specified by choosing one of the corresponding Cases 1-8 described above using the argument \code{error.type = c(1, 2, 3, 4, 5, 6, 7, 8)}.

In presence of serial correlations (cases 5 and 8), the kernel weights required for estimating the long-run covariance can be externally specified by giving a vector of weights in the argument \code{kernel.weights}. By default, the function uses internally the linearly decreasing weights of \Citet{Newey1987} and a truncation at $\left\lfloor \min\{\sqrt{n},\sqrt{T}\} \right\rfloor$. If case 7 or 8 is chosen, the method \code{summary()} calculates the realization of the bias corrected estimators and gives appropriate inferences. The bias corrected coefficients can be called by using the method \code{coef()} to the object produced by \code{summary()}. 
\begin{Schunk}
\begin{Sinput}
R> summary(Cigar.Eup)
\end{Sinput}
\begin{Soutput}
Call:
Eup.default(formula = d.l.Consumption ~ -1 + d.l.Price + d.l.Income, 
    dim.criterion = "PC3")

Residuals:
      Min        1Q    Median        3Q       Max 
-0.147000 -0.013700  0.000889  0.014100  0.093300 


 Slope-Coefficients:
           Estimate Std.Err Z value    Pr(>z)    
d.l.Price   -0.3140  0.0227  -13.90 < 2.2e-16 ***
d.l.Income   0.1590  0.0358    4.45  8.39e-06 ***
---
Signif. codes:  0 '***' 0.001 '**' 0.01 '*' 0.05 '.' 0.1 ' ' 1

Additive Effects Type:  none  

Dimension of the Unobserved Factors: 5  

Residual standard error: 0.02804 on 957 degrees of freedom,  
R-squared: 0.7033 
\end{Soutput}
\end{Schunk}

The summary output reports that \code{"PC3"} detects $5$ common factors. The effect of the differenced log-real prices for cigarettes on the differenced log-sales is negative and amounts to $-0.31$. The estimated effect of the differenced real disposable log-income per capita is $0.16$.

\begin{figure}[h]
    \centering
    \includegraphics[width=1\textwidth,height= 0.5\textwidth]{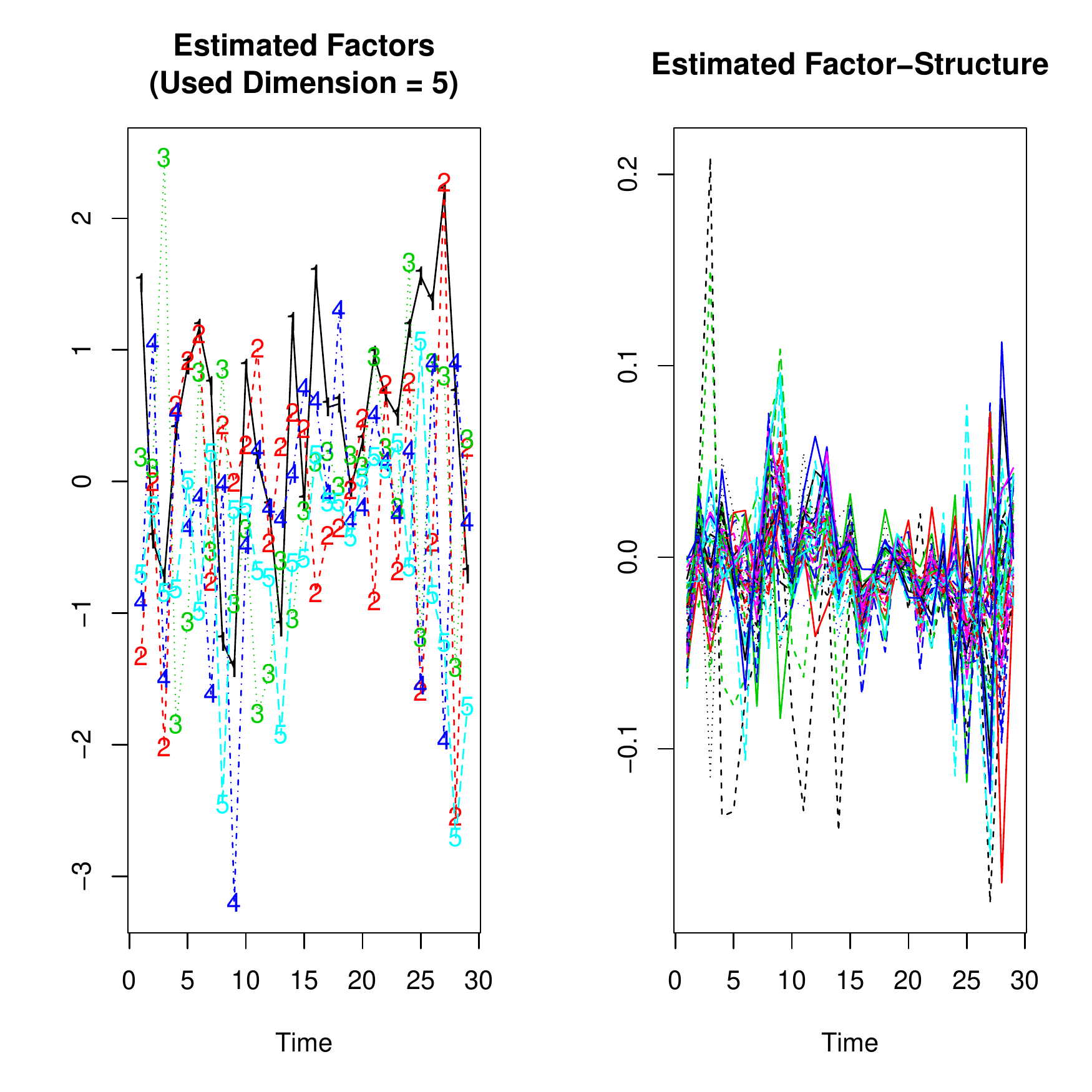} 
    \caption{{\sc Left Panel:} Estimated factors $\hat{f}_{1t},\dots,\hat{f}_{7t}$. {\sc Right panel:} Estimated time-varying individual effects $\hat{v}_{1t},\dots,\hat{v}_{nt}$.}  
  \label{EupPlot} 
\end{figure}

The estimated factors $\hat{f}_{tl}$ as well as the individual effects $\hat{v}_{it}$ can be plotted using the \code{plot()}-method for \code{summary.Eup}-objects. The corresponding graphics are shown in Figure~\ref{EupPlot}.

\begin{Schunk}
\begin{Sinput}
R> plot(summary(Cigar.Eup))
\end{Sinput}
\end{Schunk}

\section{Models with additive and interactive unobserved effects}\label{mbshtve}
Even though the classical additive \code{"individual"}, \code{"time"}, and \code{"twoways"} effects can be absorbed by the factor structure, there are good reasons to model them explicitly. On the one hand, if there are such effects in the true model, then neglecting them will result in non-efficient estimators; see \Citet{Bai2009a}. On the other hand, additive effects can be very useful for interpretation.

Consider now the following model:
\begin{equation}\label{extendmodeladd}
y_{it} = \mu + \alpha_i + \theta_t + x_{it}^{\top}\beta + \nu_{it} + \epsilon_{it}
\end{equation}
with
\begin{equation*}
\nu_{it}=\left\{
\begin{array}{ccll}
v_{it}   &=&\sum_{l =1}^d \lambda_{il}f_{lt},  &\textrm{for the model of \Citet{Bai2009a},}\\
v_{i}(t) &=&\sum_{l =1}^d \lambda_{il}f_{l}(t),&\textrm{for the model of \Citet{Kneip2009}},
\end{array}\right.
\end{equation*}
where $\alpha_i$ are time-constant individual effects and $\theta_t$ is a common time-varying effect.

In order to ensure identification of the additional additive effects  $\alpha_i$ and $\theta_t$, we need the following further restrictions: 
\begin{itemize}
\item [(d)] $\sum_{i=1}^n\lambda_{il} = 0$\quad for all \quad $l \in\{ 1, \ldots, d\}$
\item [(e)] $\sum_{t=1}^T f_{lt} = 0$\quad  for all \quad $l \in \{1, \ldots, d\}$
\item [(f)] $\sum_{i=1}^n\alpha_i = 0$ 
\item [(g)] $\sum_{t=1}^T\theta_t = 0$
\end{itemize}
By using the classical within-transformations on the observed variables, we can eliminate the additive effects $\alpha_i$ and $\theta_t$, such that
\begin{equation*}
\dot{y}_{it} =  \dot{x}_{it}^{\top}\beta + \nu_{it} + \dot{\epsilon}_{it},
\end{equation*}
where $\dot{y}_{it} =  y_{it} - \frac{1}{T}\sum_{t=1}^T y_{it} - \frac{1}{n}\sum_{i=1}^n y_{it}  + \frac{1}{nT}\sum_{t=1}^T\sum_{i=1}^n y_{it}$, $\dot{x}_{it} =  x_{it}  - \frac{1}{T}\sum_{t=1}^T x_{it} - \frac{1}{n}\sum_{i=1}^n x_{it}  + \frac{1}{nT}\sum_{t=1}^T\sum_{i=1}^n x_{it}$, and $\dot{\epsilon}_{it}  =  \epsilon_{it}  - \frac{1}{T}\sum_{t=1}^T \epsilon_{it} - \frac{1}{n}\sum_{i=1}^n \epsilon_{it}  + \frac{1}{nT}\sum_{t=1}^T\sum_{i=1}^n \epsilon_{it}$. 

Note that Restrictions (d) and (e) ensure that the transformation does not affect the time-varying individual effects $\nu_{it}$. The parameters $\mu, \alpha_i$ and $\theta_t$ can be easily estimated in a second step once an estimate of $\beta$ is obtained. Because of Restrictions (d) and (e), the solution has the same form as the classical fixed effects model. 

The parameters $\beta$ and $\nu_{it}$ can be estimated by the above introduced estimation procedures. All possible variants of model \eqref{extendmodeladd} are implemented in the functions \code{KSS()} and \code{Eup()}. The appropriate model can be specified by the argument \code{additive.effects = c("none", "individual", "time", "twoways")}: 
\begin{eqnarray*}
\textrm{\code{"none"}} &&       y_{it} = \mu + x_{it}^{\top}\beta + \nu_{it} + \epsilon_{it}\\ 
\textrm{\code{"individual"}} && y_{it} = \mu + \alpha_i + x_{it}^{\top}\beta + \nu_{it} + \epsilon_{it}\\
\textrm{\code{"time"}}&&       y_{it} = \mu + \theta_t + x_{it}^{\top}\beta + \nu_{it} + \epsilon_{it}\\
\textrm{\code{"twoways"}}&&     y_{it} = \mu + \alpha_i + \theta_t + x_{it}^{\top}\beta + \nu_{it} + \epsilon_{it}.
\end{eqnarray*}
The presence of $\mu$ can be controlled by \code{-1} in the \code{formula}-object: a formula with  \code{-1} refers to a model without intercept. However, for identification purposes, if a \code{twoways} model is specified, the presence \code{-1} in the \code{formula} will be ignored.

As an illustration, we continue with the application of the \code{KSS()}-function in Section~\ref{KSS}. The left panel of Figure~\ref{KSSM1_Fig} shows that the first common factor is nearly time-invariant. This motivates us to augment the model~\eqref{KSSM1} for a time-constant additive effects $\alpha_i$. In this case, it is convenient to use an intercept $\mu$, which yields the following model:
\begin{eqnarray}\label{KSSM2}
  \ln(\texttt{Consumption}_{it})&=&\mu+\beta_1\ln(\texttt{Price}_{it})+\beta_2\ln(\texttt{Income}_{it})+\alpha_i+ v_{i}(t)+\varepsilon_{it},\\
  \textrm{where}\quad v_{i}(t)&=&\sum_{l=1}^d\lambda_{il}\,f_l(t)\notag.
\end{eqnarray}

The estimation of the augmented model~\eqref{KSSM2} can be done using the following lines of code.

\begin{Schunk}
\begin{Sinput}
R> Cigar2.KSS <- KSS(formula = l.Consumption ~ l.Price + l.Income,
+                   additive.effects = "individual") 
R> (Cigar2.KSS.summary <- summary(Cigar2.KSS))
\end{Sinput}
\end{Schunk}

\begin{CodeChunk}
\begin{CodeOutput}
Call:
KSS.default(formula = l.Consumption ~ l.Price + l.Income, 
            additive.effects = "individual")

Residuals:
   Min     1Q Median     3Q    Max 
 -0.11  -0.01   0.00   0.01   0.12


 Slope-Coefficients:
            Estimate  StdErr z.value    Pr(>z)    
(Intercept)   4.0500  0.1760   23.10 < 2.2e-16 ***
l.Price      -0.2600  0.0222  -11.70 < 2.2e-16 ***
l.Income      0.1570  0.0381    4.11  3.88e-05 ***
---
Signif. codes:  0 '***' 0.001 '**' 0.01 '*' 0.05 '.' 0.1 ' ' 1

Additive Effects Type:  individual  

Used Dimension of the Unobserved Factors: 5  

Residual standard error: 0.000734 on 951 degrees of freedom 
R-squared: 0.99 
\end{CodeOutput} 
\end{CodeChunk}

Again, the \code{plot()} method provides a useful visualization of the results.
\begin{Schunk}
\begin{Sinput}
R> plot(Cigar2.KSS.summary)
\end{Sinput}
\end{Schunk}

\begin{figure}[htbp]
    \centering
    \includegraphics[width=1\textwidth,height=0.5\textwidth]{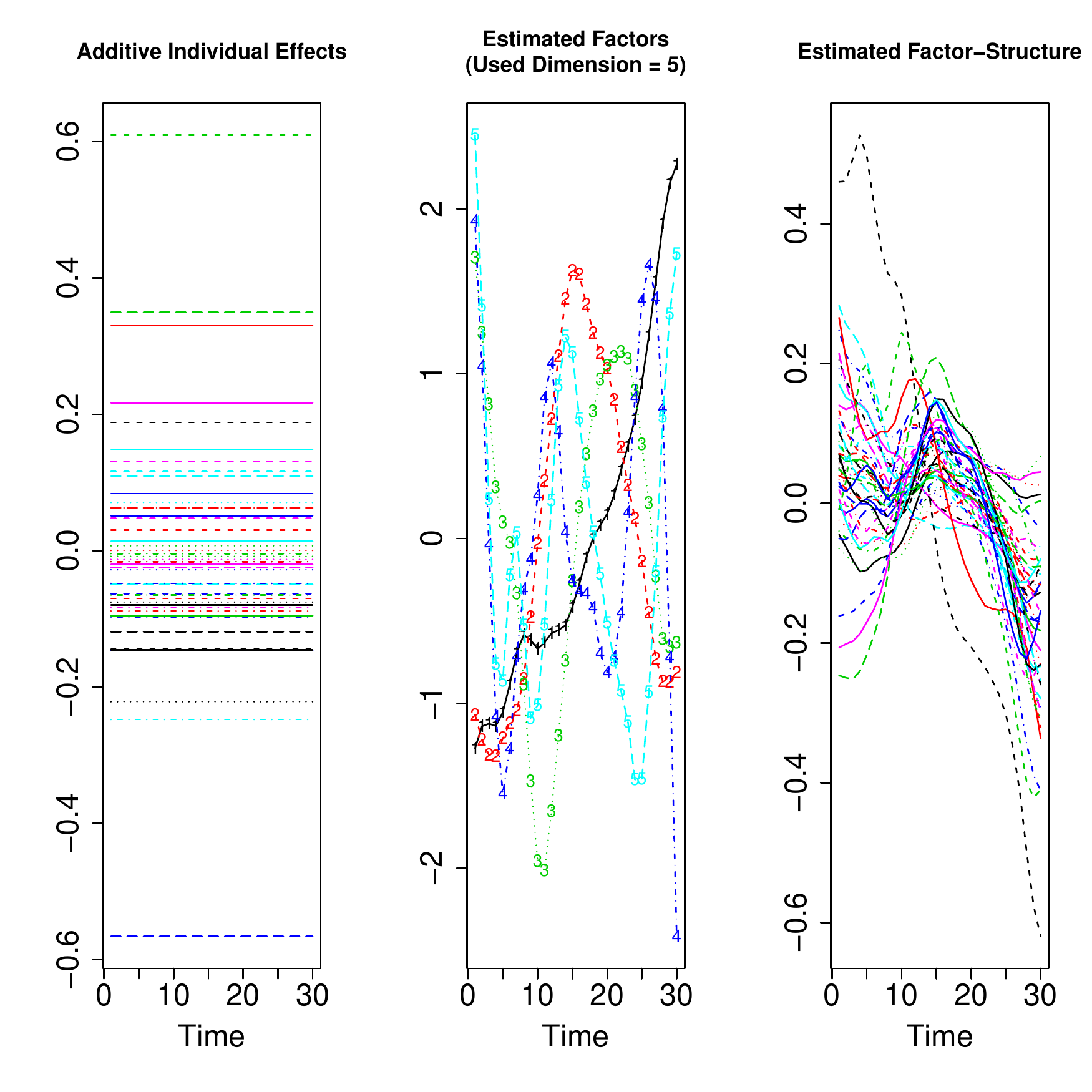} 
    \caption{{\sc Left Panel:} Estimated time-constant state-specific effects $\hat\alpha_1,\dots,\hat\alpha_n$. {\sc Middle Panel:} Estimated common factors $\hat{f}_1(t),\dots,\hat{f}_5(t)$. {\sc Right Panel:} Estimated time-varying individual effects $\hat{v}_1(t),\dots,\hat{v}_n(t)$.}  
  \label{KSSM2_Fig} 
\end{figure}

The \code{"individual"}-transformation of the data does not affect the estimation of the slope parameters, but reduces the estimated dimension from $\hat{d}=6$ to $\hat{d}=5$. The remaining five common factors $\hat{f}_1,\dots,\hat{f}_5$ correspond to those of model~\eqref{KSSM1}; see the middle panel of Figure~\ref{KSSM2_Fig}. The estimated time-constant state-specific effects $\alpha_i$ are shown in the left plot of Figure~\ref{KSSM2_Fig}. The extraction of the $\alpha_i$'s from the factor structure yields a denser set of time-varying individual effects $\hat{v}_i$ shown in the right panel of Figure~\ref{KSSM2_Fig}.

\subsection{Specification tests}\label{ADDvsINT}
Model specification is an important step for any empirical analysis. The \pkg{phtt} package is equipped with two types of specification tests: the first is a Hausman-type test appropriate for the model of \Citet{Bai2009a}; see Section~\ref{tavie}. The second one examines the existence of a factor structure in Bai's model as well as in the model of \Citet{Kneip2009}; see Section~\ref{teie}. 

\subsubsection{Testing the sufficiency of classical additive effects}\label{tavie}
For the case in which the estimated number of factors amounts to one or two ($1\leq \hat d \leq 2$), it is interesting to check whether or not these factors can be interpreted as classical \code{"individual"}, \code{"time"}, or \code{"twoways"} effects. \Citet{Bai2009a} considers the following testing problem: 
\begin{center}
  \begin{tabular}{ll}
    $H_0$: & $v_{it} =  \alpha_i + \theta_t $ \\
    $H_1$: & $v_{it} =  \sum_{l=1}^2 \lambda_{il}f_{lt}$\\
  \end{tabular}
\end{center}
The model with factor structure, as described in Section~\ref{Eup}, is consistent under both hypotheses. However, it is less efficient under $H_0$ than the classical within estimator, while the latter is inconsistent under $H_1$ if $x_{it}$ and $v_{it}$ are correlated. These conditions are favorable for applying the Hausman test:
\begin{equation}\label{HausmanTest} 
J_{Bai} = nT \left(\hat{\beta} - \hat{\beta}_{within}\right) \Delta^{-1}\left(\hat{\beta} - \hat{\beta}_{within}\right) \stackrel{a}{\sim}\chi^2_P,
\end{equation}
where $\hat{\beta}_{within}$ is the classical within least squares estimator, $ \Delta$ is the asymptotic variance of $\sqrt{nT} \left(\hat{\beta} - \hat{\beta}_{within}\right)$, $P$ is the vector-dimension of $\beta$, and $\chi^2_P$ is the $\chi^2$-distribution with $P$ degrees of freedom. 

The null hypothesis $H_0$ can be rejected, if $J_{Bai} > \chi^2_{P, 1-\alpha}$, where $\chi^2_{P, 1-\alpha}$ is the $(1-\alpha)$-quantile of the $\chi^2$ distribution with $P$ degrees of freedom.

Under i.i.d.~errors, $J_{Bai}$ can be calculated by replacing $ \Delta $ with its consistent estimator 
\begin{equation}\label{DelatEstim}
\hat{\Delta} = \left(\left(\frac{1}{nT}\sum_{i=1}^n Z_i^{\top} Z_i\right)^{-1}  -  \left(\frac{1}{nT}\sum_{i=1}^n \sum_{t=1}^T \dot{x}^{\top}_i \dot{x}_i\right)^{-1}\right)\hat{\sigma}^2,
\end{equation}
where
\begin{equation}\label{sig2s6}
\hat{\sigma}^2 = \frac{1}{nT - (n+T)\hat{d} - P +1} \sum_{i=1}^n \sum_{t=1}^T (y_{it} - x_{it}^{\top}\hat{\beta} - \sum_{l=1}^{\hat d} \hat\lambda_{il}\hat f_{lt} )^2. 
\end{equation}
The used residual variance estimator $\hat{\sigma}^2$ is chosen here, since it is supposed to be consistent under the null as well as the alternative hypothesis. The idea behind this trick is to avoid negative definiteness of $\hat{\Delta}$. But notice that even with using this construction, the possibility of getting a negative  definite variance estimator cannot be excluded. As an illustration, consider the case in which the true number of factors is greater than the number of factors used under the alternative hypothesis, i.e., the true $d > 2$. In such a case, the favorable conditions for applying the test can be violated, since the iterated least squares estimator $\hat \beta$ is computed with $\hat d \leq2$ and can be inconsistent under both hypothesis. To avoid such a scenario, we recommended to the user to calculate $\hat \beta$ with a large dimension $d_{max}$ instead of $\hat d \leq2$.

The test is implemented in the function \code{checkSpecif()}, which takes the following arguments: 
\begin{Schunk}
\begin{Sinput}
R> checkSpecif(obj1, obj2, level = 0.05)
\end{Sinput}
\end{Schunk}

The argument \code{level} is used to specify the significance level. The arguments \code{obj1} and \code{obj2} take both objects of class \code{Eup} produced by the function \code{Eup()}: 
\begin{itemize}
\item[\code{obj1}] Takes an \code{Eup}-object from an estimation with \code{"individual"}, \code{"time"}, or \code{"twoways"} effects and a factor dimension equal to $d=0$; specified as \code{factor.dim = 0}. 
\item[\code{obj2}] Takes an \code{Eup}-object from an estimation with \code{"none"}-effects and a large factor dimension $d_{max}$; specified with the argument \code{factor.dim}.\\
\end{itemize}
If the test statistic is negative (due to the negative definiteness of $\hat{\Delta}$), the \code{checkSpecif()} prints an error message.

\begin{Schunk}
\begin{Sinput}
R> twoways.obj     <- Eup(d.l.Consumption ~  -1 + d.l.Price + d.l.Income, 
+                        factor.dim = 0, additive.effects = "twoways")
R> not.twoways.obj <- Eup(d.l.Consumption ~  -1 + d.l.Price + d.l.Income, 
+                        factor.dim = 2, additive.effects = "none")  
R> checkSpecif(obj1 = twoways.obj, obj2 = not.twoways.obj, level = 0.01)
\end{Sinput}
\end{Schunk}

\begin{CodeChunk}
\begin{CodeOutput}
Error in checkSpecif(obj1 = twoways.obj, obj2 = not.twoways.obj, 
level = 0.01): 
The assumptions of the test are not fulfilled.
The (unobserved) true number of factors is probably greater than 2.
\end{CodeOutput}
\end{CodeChunk}

Notice that the Hausman test of \Citet{Bai2009a} assumes the within estimator to be inconsistent under the alternative hypothesis, which requires $x_{it}$ to be correlated with $v_{it}$. If this assumption is violated, the test can suffer from power to reject the null hypothesis, since the within estimator becomes consistent under both hypothesis. 

\Citet{Bai2009a} discusses in his supplementary material another way to check whether a classical panel data with fixed additive effects is sufficient to describe the data. His idea consists of estimating the factor dimension after eliminating the additive effects as described in Section \ref{mbshtve}. If the obtained estimate of $d$ is zero, the additive model can be considered as a reasonable alternative for the model with factor structure. But note that this procedure can not be considered as a formal testing procedure, since information about the significance level of the decision are not provided.  

An alternative test for the sufficiency of a classical additive effects model can be given by manipulating the test proposed by \Citet{Kneip2009} as described in the following section.

\subsubsection{Testing the existence of common factors}\label{teie}
This section is concerned with testing the existence of common factors. In contrast to the Hausman type statistic discussed above, the goal of this test is not merely to decide which model specification is more appropriate for the data, but rather to test in general the existence of common factors beyond the possible presence of additional classical \code{"individual"}, \code{"time"}, or \code{"twoways"} effects in the model. 

This test relies on using the dimensionality criterion proposed by \Citet{Kneip2009} to test the following hypothesis after eliminating eventual additive \code{"individual"}, \code{"time"}, or \code{"twoways"} effects:
\begin{center}
  \begin{tabular}{ll}
    $H_0$: & $d = 0$\\
    $H_1$: & $d > 0$\\
  \end{tabular}
\end{center}
Under $H_0$ the slope parameters $\beta$ can be estimated by the classical within estimation method. In this simple case, the dimensionality test of \Citet{Kneip2009} can be reduced to the following test statistic:
\begin{equation*}
  J_{KSS} = \frac{n \; tr(\hat \Sigma_w) - (n-1)(T-1)\hat{\sigma}^2 }{\sqrt{2n}(T-1) \hat{\sigma}^2 } \stackrel{a}{\sim} N(0,1),
\end{equation*}
where $\hat{\Sigma}_w$ is the covariance matrix of the within residuals. The reason for this simplification is that under $H_0$ there is no need for smoothing, which allows us to set $\kappa = 0$. 
                
We reject $H_0\textrm{: }d=0$ at a significance level $\alpha$, if $J_{KSS} > z_{1 - \alpha}$, where $z_{1 - \alpha}$ is the $(1-\alpha)$-quantile of the standard normal distribution. It is important to note that the performance of the test depends heavily on the accuracy of the variance estimator $\hat{\sigma}^2$. We propose to use the variance estimators~\eqref{sig2s2} or \eqref{sig2s6}, which are consistent under both hypotheses as long as $\hat{d}$ is greater than the unknown dimension $d$. Internally, the test procedure sets $\hat{d}=$\code{d.max} and $\sigma^2$ as in \eqref{sig2s6}.

This test can be performed for \code{Eup}- as well as for \code{KSS}-objects by using the function \code{checkSpecif()} leaving the second argument \code{obj2} unspecified. In the following, we apply the test for both models:  

For the model of \Citet{Bai2009a}:
\begin{Schunk}
\begin{Sinput}
R> Eup.obj <- Eup(d.l.Consumption ~  -1 + d.l.Price + d.l.Income, 
+                additive.effects = "twoways")
R> checkSpecif(Eup.obj, level = 0.01)
\end{Sinput}
\begin{Soutput}
----------------------------------------------
Testing the Presence of Interactive Effects
Test of Kneip, Sickles, and Song (2012)
----------------------------------------------
H0: The factor dimension is equal to 0.

Test-Statistic        p-value    crit.-value     sig.-level 
         13.29           0.00           2.33           0.01 
\end{Soutput}
\end{Schunk}

For the model of \Citet{Kneip2009}:
\begin{Schunk}
\begin{Sinput}
R> KSS.obj <- KSS(l.Consumption ~  -1 + l.Price + l.Income, 
+                additive.effects = "twoways")
R> checkSpecif(KSS.obj, level = 0.01)
\end{Sinput}
\begin{Soutput}
----------------------------------------------
Testing the Presence of Interactive Effects
Test of Kneip, Sickles, and Song (2012)
----------------------------------------------
H0: The factor dimension is equal to 0.

Test-Statistic        p-value    crit.-value     sig.-level 
     104229.55           0.00           2.33           0.01 
\end{Soutput}
\end{Schunk}

The null hypothesis $H_0\textrm{: }d=0$ can be rejected for both models at a significance level $\alpha=0.01$.

\section{Interpretation}
This section is intended to outline an exemplary interpretation of the panel model~\eqref{KSSM2}, which is estimated by the function \code{KSS()} in Section~\ref{mbshtve}. The interpretation of models estimated by the function \code{Eup()} can be done accordingly. For convenience sake, we re-write the model~\eqref{KSSM2} in the following:
\begin{eqnarray*}
  \ln(\texttt{Consumption}_{it})&=&\mu+\beta_1\ln(\texttt{Price}_{it})+\beta_2\ln(\texttt{Income}_{it})+\alpha_i+ v_{i}(t)+\varepsilon_{it},\\
  \textrm{where}\quad v_{i}(t)&=&\sum_{l=1}^d\lambda_{il}\,f_l(t)\notag.
\end{eqnarray*}

A researcher, who chooses the panel models proposed by \Citet{Kneip2009} or \Citet{Bai2009a}, will probably find them attractive due to their ability to control for very general forms of unobserved heterogeneity. Beyond this, a further great advantage of these models is that the time-varying individual effects $v_{i}(t)$ provide a valuable source of information about the \emph{differences} between the individuals $i$. These differences are often of particular interest as, e.g., in the literature on stochastic frontier analysis. 

The left panel of Figure~\ref{KSSM2_Fig} shows that the different states $i$ have considerable different time-constant levels $\hat\alpha_i$ of cigarette consumption. A classical further econometric analysis could be to regress the additive individual effects $\hat\alpha_i$ on other time-constant variables, such as the general populations compositions, the cigarette taxes, etc. 

The right panel of Figure~\ref{KSSM2_Fig} shows the five estimated common factors $\hat{f}_1(t),\dots,\hat{f}_5(t)$. It is a good practice to start the interpretation of the single common factors with an overview about their importance in describing the differences between the $v_i(t)$'s, which is reflected in the variances of the individual loadings parameters $\hat\lambda_{il}$. A convenient depiction is the quantity of variance-shares of the individual loadings parameters on the total variance of the loadings parameters 
$$\textrm{\code{coef(Cigar2.KSS)\$Var.shares.of.loadings.param[$l$]}}=\textrm{V}(\hat\lambda_{il})/\sum_{k=1}^{\hat{d}}\textrm{V}(\hat\lambda_{ik}),$$
which is shown for all common functions $\hat{f}_1(t),\dots,\hat{f}_5(t)$ in the following table:
\begin{table}[ht]
  \centering
  \begin{tabular}{c|l}
    Common Factor &Share of total variance of $v_i(t)$\\
    \hline
    $\hat{f}_1(t)$&\code{coef(Cigar2.KSS)\$Var.shares.of.loadings.param[1] }$=66.32\%$\\
    $\hat{f}_2(t)$&\code{coef(Cigar2.KSS)\$Var.shares.of.loadings.param[2] }$=24.28\%$\\
    $\hat{f}_3(t)$&\code{coef(Cigar2.KSS)\$Var.shares.of.loadings.param[3] }$= 5.98\%$\\
    $\hat{f}_4(t)$&\code{coef(Cigar2.KSS)\$Var.shares.of.loadings.param[4] }$= 1.92\%$\\
    $\hat{f}_5(t)$&\code{coef(Cigar2.KSS)\$Var.shares.of.loadings.param[5] }$= 1.50\%$\\
  \end{tabular}
  \caption{List of the variance shares of the common factors $\hat{f}_1(t),\dots,\hat{f}_5(t)$.}
  \label{Tab1}
\end{table}

The values in Table~\ref{Tab1} suggest to focus on the first two common factors, which explain together about $90\%$ of the total variance of the time-varying individual effects $\hat{v}_i(t)$.

\begin{figure}[h]
    \centering
    \includegraphics[width=1\textwidth,height=.7\textwidth]{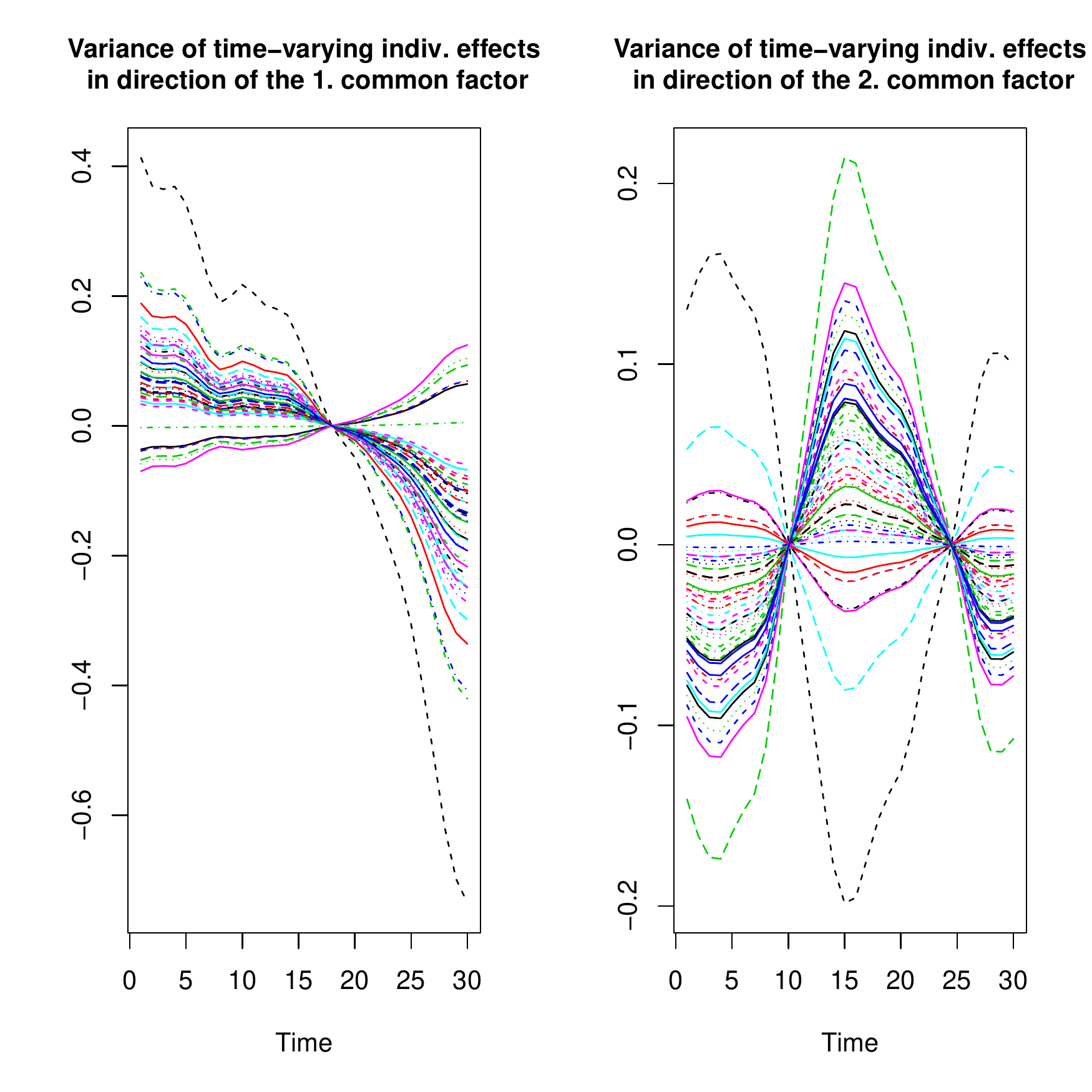} 
    \caption{{\sc Left Panel: }Visualization of the differences of the time-varying individual effects $v_i(t)$ in the direction of the first factor $\hat{f}_1(t)$ (i.e., $\hat\lambda_{i1}\hat{f}_1(t)$). {\sc Right Panel: }Visualization of the differences of the time-varying individual effects $v_i(t)$ in the direction of the second factor $\hat{f}_2(t)$ (i.e., $\hat\lambda_{i2}\hat{f}_2(t)$).}  
  \label{Factor2_Fig} 
\end{figure}

The first two common factors 
\begin{center}
\begin{tabular}{rl}
  $\textrm{\code{coef(Cigar2.KSS)\$Common.factors[,$1$]}}=\hat{f}_{1}(t)$& and\\
  $\textrm{\code{coef(Cigar2.KSS)\$Common.factors[,$2$]}}=\hat{f}_{2}(t)$&    \\ 
\end{tabular}
\end{center}
are plotted as black and red lines in the middle panel of Figure~\ref{KSSM2_Fig}. Figure~\ref{Factor2_Fig} visualizes the differences of the time-varying individual effects $v_i(t)$ in the direction of the first common factor (i.e., $\hat\lambda_{i1}\hat{f}_1(t)$) and in the direction of the second common factor (i.e., $\hat\lambda_{i2}\hat{f}_2(t)$). As for the time-constant individual effects $\hat\alpha_i$ a further econometric analysis could be to regress the individual loadings parameters $\hat\lambda_{i1}$ and $\hat\lambda_{i2}$ on other explanatory time-constant variables.

Generally, for both models proposed by \Citet{Kneip2009} and \Citet{Bai2009a} the time-vaying individual effects 
$$\nu_{it}=\sum_{l=1}^d\lambda_{il}f_{lt}$$ 
can be interpreted as it is usually done in the literature on factor models. An important topic that is not covered in this section is the rotation of the common factors. Often, the common factors $f_l$ can be interpreted economically only after the application of an appropriate rotation scheme for the set of factors $\hat{f}_1,\dots,\hat{f}_{\hat{d}}$. The latter can be done, e.g., using the function \code{varimax()} from the \pkg{stats} package. Alternatively, many other rotation schemes can be found in the \pkg{GPArotation} package (\cite{stats.pkg}, \cite{GPA.pkg}). Sometimes, it is also preferable to standardize the individual loadings parameters instead of the common factors as it is done, e.g., in \Citet{ahn2001gmm}. This can be done by choosing \code{restrict.mode = c("restrict.loadings")} in the functions \code{KSS()} and \code{Eup()} respectively. 

\section{Summary}
This paper introduces the \proglang{R} package \pkg{phtt} for the new class of panel models proposed by \Citet{Bai2009a} and \Citet{Kneip2009}. The two main functions of the package are the \code{Eup()}-function for the estimation procedure proposed in \Citet{Bai2009a} and the \code{KSS()}-function for the estimation procedure proposed in \Citet{Kneip2009}. Both of the main functions are supported by the usual \code{print()}-, \code{summary()}-, \code{plot()}-, \code{coef()}- and \code{residuals()}-methods. While parts of the method of \Citet{Bai2009a} are available for commercially available software packages, the estimation procedure proposed by \Citet{Kneip2009} is not available elsewhere. A further remarkable feature of our \pkg{phtt} package is the \code{OptDim()}-function, which provides an ease access to many different dimensionality criteria proposed in the literature on factor models. The usage of the functions is demonstrated by a real data application. 

\section{Acknowledgment}
The authors wish to thank the referees for their many helpful comments and suggestions that greatly improved the paper. The work of Dominik Liebl was supported from the IAP Research NetworkP7/06 of the Belgian State (Belgian Science Policy). 
\newpage

\end{document}